\documentclass{aa}  

\usepackage{graphicx}
\usepackage{txfonts}
\usepackage{longtable}
\usepackage{rotating}
\usepackage{natbib}
\usepackage{graphics}
\usepackage{psfrag}
\usepackage{amssymb}
\usepackage{xspace}
\usepackage{color}
\usepackage{mathtools}
\usepackage{amsmath}
\bibpunct{(}{)}{;}{a}{}{,}
\def\hh{{\rm H_2}}

\def\nhhh{n_{\rm H_3^+}}

\def\hho{H$_{\rm 2}$O}
\def\co{CO}
\def\oh{OH}
\def\vo{VO}
\def\sio{SiO}
\def\tio{TiO}

\def\coo{CO$_{\rm 2}$}

\def\chhhh{CH$_{\rm 4}$}
\def\nhhh{NH$_{\rm 3}$}
\def\hcn{HCN}
\def\hh{H$_{\rm 2}$}

\def\hminus{H$^-$}
\def\heminus{He$^-$}

\begin{document}

\title{Non-local thermodynamic equilibrium effects determine the upper atmospheric temperature structure of the ultra-hot Jupiter KELT-9b}
\subtitle{}

\titlerunning{Non-local thermodynamic equilibrium effects in the upper atmosphere of KELT-9b}
\authorrunning{Fossati et al.}

\author{L. Fossati\inst{1}	\and
	M. E. Young\inst{2,1}	\and
	D. Shulyak\inst{3}	\and
	T. Koskinen\inst{4}	\and
	C. Huang\inst{4}	\and
	P. E. Cubillos\inst{1}	\and
	K. France\inst{5,6}	\and
	A. G. Sreejith\inst{1}
	}
\institute{
Space Research Institute, Austrian Academy of Sciences, Schmiedlstrasse 6, A-8042 Graz, Austria\\
\email{luca.fossati@oeaw.ac.at}
\and 
Department of Physics, University of Oxford, Denys Wilkinson Building, Keble Road, Oxford, OX1 3RH United Kingdom
\and 
Max-Planck Institut f\"ur Sonnensystemforschung, Justus-von-Liebig-Weg 3, D-37077, G\"ottingen, Germany
\and
Lunar and Planetary Laboratory, University of Arizona, 1629 East University Boulevard, Tucson, AZ 85721-0092, USA
\and
Laboratory for Atmospheric and Space Physics, University of Colorado, 600 UCB, Boulder, CO 80309, USA
\and
Center for Astrophysics and Space Astronomy, University of Colorado, 389 UCB, Boulder, CO 80309, USA
}

\date{}

\abstract
{Several observational and theoretical results indicate that the atmospheric temperature of the ultra-hot Jupiter KELT-9b in the main line formation region is a few thousand degrees higher than predicted by self-consistent models.}
{We aim to test whether non-local thermodynamic equilibrium (NLTE) effects are responsible for the presumably higher temperature.}
{We employ the Cloudy NLTE radiative transfer code to self-consistently compute the upper atmospheric temperature-pressure (TP) profile of KELT-9b, assuming solar metallicity and accounting for Roche potential. In the lower atmosphere, we use an updated version of the HELIOS radiative-convective equilibrium code to constrain the Cloudy model.}
{The Cloudy NLTE TP profile is $\approx$2000\,K hotter than that obtained with previous models assuming local thermodynamic equilibrium (LTE). In particular, in the 1--10$^{-7}$\,bar range the temperature increases from $\approx$4000\,K to $\approx$8500\,K, remaining roughly constant at lower pressures. We find that the high temperature in the upper atmosphere of KELT-9b is driven principally by NLTE effects modifying the Fe and Mg level populations, which strongly influence the atmospheric thermal balance. We employ Cloudy to compute LTE and NLTE synthetic transmission spectra on the basis of the TP profiles computed in LTE and NLTE, respectively, finding that the NLTE model generally produces stronger absorption lines than the LTE model (up to 30\%), which is largest in the ultraviolet. We compare the NLTE synthetic transmission spectrum with the observed H$\alpha$ and H$\beta$ line profiles obtaining an excellent match, thus supporting our results.}
{The NLTE synthetic transmission spectrum can be used to guide future observations aiming at detecting features in the planet's transmission spectrum. Metals, such as Mg and Fe, and NLTE effects shape the upper atmospheric temperature structure of KELT-9b and thus affect the mass-loss rates derived from it. Finally, our results call for checking whether this is the case also of cooler planets.}

\maketitle
%
%
\section{Introduction} \label{sec:introduction}
Ultra-hot Jupiters, that is planets for which the continuum is dominated by H$^-$ and the optical and infrared spectral ranges are dominated by metal absorption and lack of molecular absorption \citep{arcangeli2018,parmentier2018,kitzmann2018,lothringer2018}, are becoming prime targets for atmospheric characterisation. This is mostly due to their high atmospheric temperature, which ensures large pressure scale heights, hence large (detectable) atmospheric signals in transmission and emission, and absence of aerosols, at least on the day-side. Furthermore, these planets typically orbit rather bright stars, favouring atmospheric characterisation observations.

About half a dozen ultra-hot Jupiters have been detected and observed to date, mostly in transmission and employing high resolution spectroscopy from the ground, though some (space-based) low resolution spectroscopic observations have also been obtained and analysed. However, the interest of the community on ultra-hot Jupiters has significantly increased following the detection of KELT-9b, also known as HD\,195689\,b, which is the hottest exoplanet orbiting a non-degenerate star known to date \citep{gaudi2017}. Both TESS and Spitzer phase curve observations have been employed to measure the day- and night-side planetary temperatures, obtaining values of roughly 4600 and 3040\,K, respectively \citep{mansfield2020,wong2020}.

KELT-9b has been one of the most targeted planets for ground-based high-resolution transmission spectroscopy observations. Both hydrogen Balmer lines and metal features have been detected in the planetary transmission spectrum \citep[e.g.][]{yan2018,hoeijmakers2018,hoeijmakers2019,yan2019, borsa2019,cauley2019,turner2020,pino2020,wyttenbach2020}. Similar results have also been obtained for a few other (not so extreme) ultra-hot Jupiters, such as WASP-33b \citep[e.g.][]{vonessen2019,yan2019,yan2020a,cauley2020,nugroho2020a}, WASP-121b \citep[e.g.][]{sing2019,gibson2020,benyami2020,bourrier2020,cabot2020, hoeijmakers2020b}, WASP-76b \citep[e.g.][]{seidel2019,ehrenreich2020,tabernero2020,edwards2020,fu2020}, WASP-189b \citep[e.g.][]{yan2020b}, and MASCARA-2b \citep[e.g.][]{casasayas2018,casasayas2019,hoeijmakers2020a, nugroho2020b,stangret2020}.

In the case of KELT-9b, the observations led to constraining the planetary atmospheric temperature-pressure (TP) profile, in addition to the composition from the detection of metal features. \citet{lothringer2018} employed the PHOENIX stellar and planetary atmosphere code to perform forward modelling of the TP profile of synthetic ultra-hot Jupiters of different temperatures assuming local thermodynamic equilibrium (LTE). They concluded that the upper atmospheric layers of these planets are characterised by a temperature inversion. According to the PHOENIX calculations, for KELT-9b this temperature inversion should lead to upper atmospheric temperatures of the order of about 6500\,K \citep[see also][]{lothringer2020}. \citet{pino2020} detected Fe{\sc i} emission from secondary eclipse ground-based high-resolution observations and arrived at the conclusion that the planetary atmosphere is indeed characterised by an inverted TP profile, confirming the modelling predictions. However, \citet{pino2020} were unable to quantify the temperature inversion. \citet{garcia2019} modelled the planetary upper atmosphere assuming a pure hydrogen composition and accounting for non-local thermodynamic equilibrium (NLTE) effects. They argued that the upper atmospheric temperature should be of the order of 15000\,K, but also mentioned that the inclusion of metals in the model would likely decrease it by 1000--2000\,K. 

The detection of the hydrogen Balmer lines provides valuable constraints on the TP profile and abundances in the upper atmosphere \citep{garcia2019,turner2020,wyttenbach2020,fossati2020}. Furthermore, this information enables one to constrain the energetics driving atmospheric escape \citep{yan2018,fossati2018,wyttenbach2020}, despite the fact that the Balmer lines do not probe the atmosphere beyond the planetary Roche lobe \citep{turner2020}. Accounting for NLTE effects, both \citet{garcia2019} and \citet{turner2020} showed that fitting the observed hydrogen Balmer lines requires an upper atmospheric temperature significantly hotter than 6500\,K, which is the value predicted by PHOENIX LTE forward models \citep{lothringer2018,fossati2018,lothringer2020}. \citet{wyttenbach2020} employed a retrieval technique to constrain the atmospheric TP profile, impact of NLTE effects, and mass-loss rate. They obtained an atmospheric temperature of 9600$\pm$1200\,K, a density of excited (i.e. in the n\,=\,2 state) hydrogen relative to the total amount of hydrogen of about 10$^{-11}$, and a mass loss rate of the order of 10$^{14}$\,g\,s$^{-1}$. However, they assumed an isothermal TP profile, which \citet{fossati2020} showed being an extreme and likely unphysical assumption. Furthermore, they considered the ratio between LTE and NLTE hydrogen densities as a free parameter, which led to significant degeneracies among the output parameters.

In an attempt to overcome the assumptions of \citet{wyttenbach2020}, \citet{fossati2020} constructed a large sample of artificial TP profiles, characterised by a temperature inversion of varying shape and location, looking for those leading to best fit the observed hydrogen Balmer lines. They accounted for NLTE effects in the spectral synthesis calculations employing the Cloudy code \citep{ferland2017}. They found that the family of TP profiles best fitting the observations is characterised by an inverted temperature profile starting at pressures higher than 10$^{-4}$\,bar and that the upper atmospheric temperature could be as high as 10000\,K. Their results also showed that the LTE assumption leads to overestimating the strength of the hydrogen Balmer lines. Driven by the significant metal ionisation in the upper atmosphere extracted from the Cloudy simulations, \citet{fossati2020} suggested that metal photoionisation might be responsible for driving the properties of the TP profile and in particular for the $\approx$3000--4000\,K difference between the TP profile best fitting the hydrogen Balmer lines and what was predicted by previous models. 

Cloudy is a code designed to simulate physical conditions within gas clouds ranging from the intergalactic medium to the high-density LTE limit \citep{ferland1998,ferland2013,ferland2017}. Within the exoplanet field, Cloudy has been employed to compute the physical conditions of planetary escaping gas interacting with the stellar radiation and wind \citep{turner2016}, to obtain heating and cooling rates to be used in exoplanetary hydrodynamic simulations \citep{salz2015,salz2016,salz2018,salz2019}, and to compute transmission spectra for a wide range of planets \citep{salz2018,salz2019,turner2020,cubillos2020,fossati2020,young2020a,young2020b}. In this work, we extend some of previous Cloudy exoplanet modelling work and in particular build up on the results of \citet{fossati2020} employing forward modelling both in LTE and NLTE to construct TP profiles of KELT-9b with the aim of exploring in detail the origin of the upper atmospheric heating. We also employ the LTE and NLTE TP profiles to generate transmission spectra and explore the impact of the LTE assumption on the interpretation of the observations. We finally compare the NLTE transmission spectrum with the observed hydrogen Balmer lines, but leave the comparison with the observed metal lines to future dedicated works.

This paper is organised as follows. Section~\ref{sec:modelling} describes the adopted modelling scheme and tools. In Section~\ref{sec:results}, we present the results obtained from forward NLTE modelling of the atmospheric structure of KELT-9b, while in Section~\ref{sec:discussion} we discuss the results and compare them with the observed H$\alpha$ and H$\beta$ line profiles. We draw the conclusions of this work in Section~\ref{sec:conclusion}.
\section{Atmospheric modelling}\label{sec:modelling}
We generated synthetic TP profiles for the ultra-hot Jupiter KELT-9b combining the results obtained employing the {\sc HELIOS} code \citep{malik2017,malik2019} for the lower atmosphere (i.e. P\,$\gtrsim$\,10$^{-4}$\,bar) and the Cloudy code \citep{ferland2017} for the upper atmosphere (i.e. P\,$\lesssim$\,10$^{-4}$\,bar). The need for combining the results of two distinct codes comes from the fact that HELIOS uses opacity tables computed under LTE assumption, which cannot be changed in the current version of the code, and NLTE effects are important in the upper atmosphere \citep[e.g.][see also below]{fossati2020}. Instead, Cloudy, which can account for NLTE effects, is unreliable for total gas densities larger than roughly 10$^{15}$\,cm$^{-3}$, hence in the lower atmosphere. This density limit is driven by the approximations involved in the treatment of three-body recombination - collisional ionisation for the heavy elements \citep[see Section 3 of][]{ferland2017}. However, tests run with Cloudy indicate that this most strongly affects calculations in which Cloudy is allowed to compute the temperature structure at high gas densities, while those with a fixed temperature profile appear to  be significantly less affected \citep{young2020a,fossati2020}. This is because the approximations impact mostly the heating and cooling functions that are involved only in calculating the temperature structure, and not the spectra.

Here below, we describe in detail the modelling conducted with {\sc HELIOS} and Cloudy. For all calculations, we considered the system parameters given by \citet{borsa2019} and listed in \citet[][Table 2]{fossati2020}. The stellar spectral energy distribution we considered as input to compute the TP profiles is that presented by \citet[][effective temperature of 10000\,K]{fossati2018}.
\subsection{Lower atmosphere: P\,$\gtrsim$\,10$^{-4}$\,bar}\label{sec:lower}
\begin{figure}[h!]
\includegraphics[width=\hsize,clip]{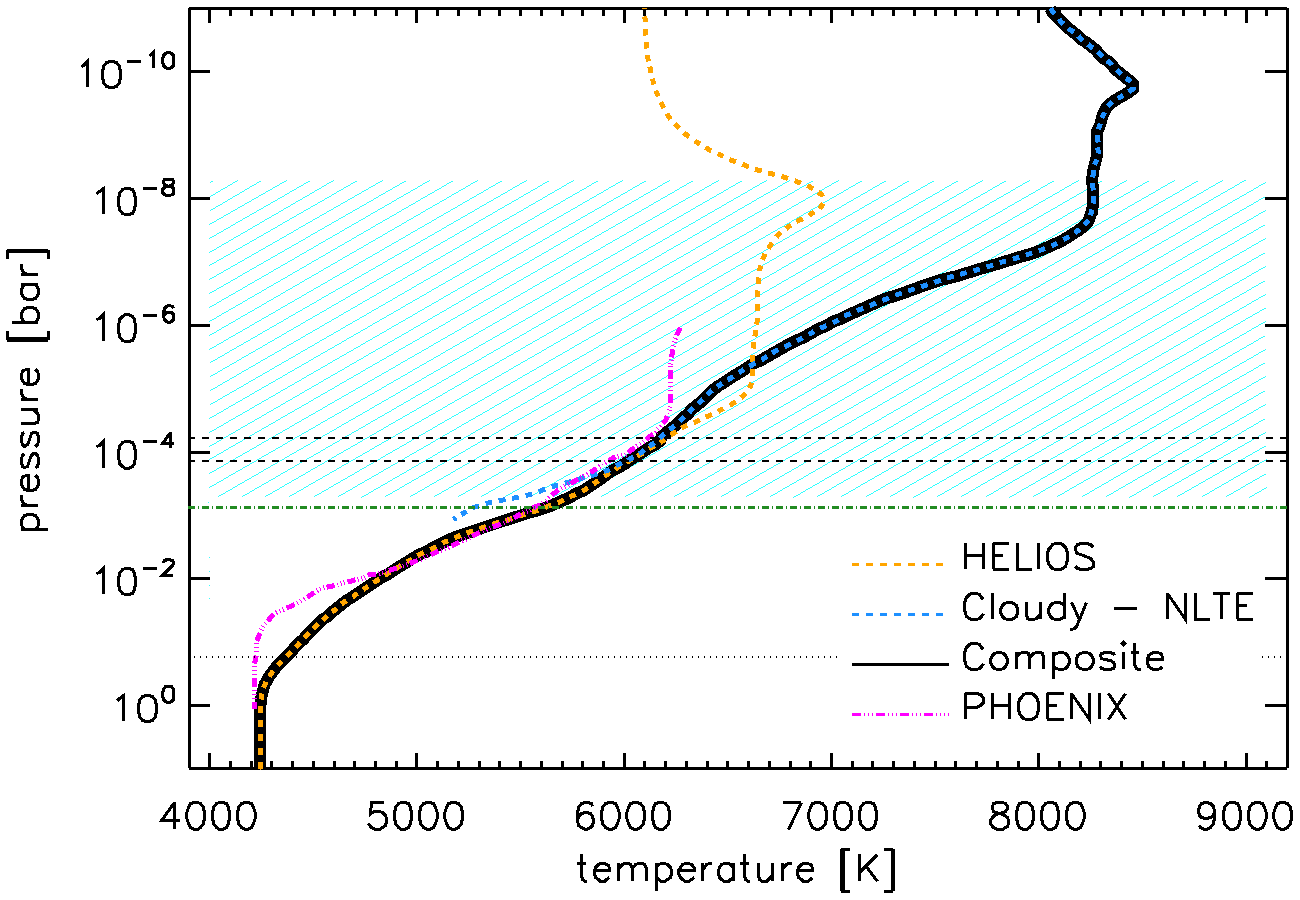}
\caption{Temperature-pressure profiles obtained employing HELIOS (orange dashed line) and Cloudy (in NLTE; bright blue dashed line). The black solid line shows the composite TP profile, while the horizontal black dashed lines mark the location within which the HELIOS and Cloudy TP profiles have been merged. The horizontal dark green dash-dotted line indicates the location of Cloudy's upper density limit (i.e. 10$^{15}$\,cm$^{-3}$). The magenta dash-dotted line shows the TP profile computed with PHOENIX \citep{lothringer2020}. The hatched area indicates the main line formation region \citep{turner2020,fossati2020}.}
\label{fig:compositeTP}
\end{figure}

To calculate the structure of the dense layers of the planetary atmosphere, we employed the \textsc{HELIOS}\footnote{\tt https://github.com/exoclime/HELIOS} radiative-convective equilibrium code \citep{malik2017,malik2019}. The inputs for the \textsc{HELIOS} code are planetary mass, radius, and atmospheric abundances, orbital semi-major axis, and stellar radius and effective temperature. Because \textsc{HELIOS} does not include photochemistry, we assumed equilibrium abundances, which we obtained with the \textsc{FastChem}\footnote{\tt https://github.com/exoclime/FastChem} code \citep{stock2018}. Chemical equilibrium is a reasonable approximation, because the concentrations of many major species in deep atmospheric layers of very hot planets are close to their chemical equilibrium values \citep[see e.g.][]{shulyak2020}. Our simulation setup includes 100 atmospheric layers distributed logarithmically between 100 and 10$^{-9}$\,bar. The equilibrium temperature of the planet is controlled by the heat redistribution parameter $f$, which accounts for the (presumably unknown) redistribution efficiency between day and night sides of the planet. By adjusting this parameter one can fine tune the desired day-side temperature of the planet. We set $f$ equal to 0.63 that is the value for which the temperature at pressures higher than 10$^{-4}$\,bar best reproduces the TP profile computed with the {\sc PHOENIX} code \citep{fossati2018,lothringer2020}. Furthermore, with this $f$ value, at pressures around 10$^{-1}$--10$^{-2}$\,bar the temperature lies between 4500 and 5000\,K, in agreement with the measured day-side temperature of about 4600\,K \citep{gaudi2017,mansfield2020}. Figure~\ref{fig:compositeTP} presents the {\sc HELIOS} TP profile, in comparison to the {\sc PHOENIX} TP profile.

The hot temperature of KELT-9b's atmosphere requires several additional opacity sources to be added compared to what present in the public version of the \textsc{HELIOS} code. First, we included atomic line opacity due to neutral and singly ionised atoms, namely C{\sc i-ii}, Cr{\sc i-ii}, Fe{\sc i-ii}, K{\sc i-ii}, Mg{\sc i-ii}, Na{\sc i-ii}, O{\sc i-ii}, and Si{\sc i-ii}. These are the elements that were found to contribute most to the line opacity throughout the planetary atmosphere. The original line lists are those produced by R.~Kurucz\footnote{\tt http://kurucz.harvard.edu/linelists/gfall/} \citep{kurucz2018}. For each atom, we generated cross sections for different sets of temperatures, pressures, and wavelengths using the \textsc{HELIOS-K}\footnote{\tt https://github.com/exoclime/HELIOS-K} package \citep{grimm2015,grimm2021}. The molecular line opacity includes such molecules as \chhhh, \coo, \co, \hho, \hcn, \nhhh, \oh, \sio, \tio, and \vo. The pretabulated cross sections of these molecules were taken from the public opacity database for exoplanetary atmospheres\footnote{\tt https://dace.unige.ch/opacity}. The generated cross sections were then weighted with the number density of the corresponding species computed under the LTE assumption and using the \textsc{ktable} code distributed together with the HELIOS installation.

Next, we extended \textsc{HELIOS} with new continuum opacity sources. These include bound-free and free-free transitions of \hminus, free-free transitions of \heminus, Rayleigh scattering on H{\sc i} and He{\sc i} atoms, and Thomson scattering on free electrons. Additionally, we included bound-free opacity for the following atoms: H{\sc i}, C{\sc i-ii}, N{\sc i-ii}, O{\sc i-ii}, Ne{\sc i-ii}, Mg{\sc i-ii}, Al{\sc i}, Si{\sc i-ii}, Ca{\sc ii}, and Fe{\sc i-ii}. Finally, we included the continuum opacity of the OH and CH molecules. All relevant numerical routines were taken from the LLmodels stellar model atmosphere code \citep[see][and references therein]{shulyak2004}. The other continuum opacities include Rayleigh scattering by \hho, \hh, and \coo\ molecules, and collisionally-induced absorption due to H$_{\rm 2}$-H$_{\rm 2}$ and H$_{\rm 2}$-He following \citet{borysow2001}, \citet{borysow2002}, and \citet{borysow1989}. This is the same set of tools that \citet{fossati2020} employed to compute synthetic hydrogen Balmer line profiles assuming LTE.
\subsection{Upper atmosphere: P\,$\lesssim$\,10$^{-4}$\,bar}\label{sec:upper}
\subsubsection{Cloudy}
We modelled the planetary upper atmosphere employing Cloudy\footnote{\tt https://gitlab.nublado.org/cloudy/cloudy/-/wikis/home} (version 17.02), which is a microphysics spectral synthesis code designed to simulate physical conditions within an astrophysical plasma, predicting the emitted/absorbed spectrum, further accounting for (photo)chemistry and NLTE effects \citep{ferland1998,ferland2013,ferland2017}. Cloudy accounts for NLTE effects by explicitly computing level populations and ionisation, without resourcing to the Boltzmann and Saha equations, and can consider non-Planck radiation field and non-Maxwellian velocity distribution. We remark that the gas in the planetary atmosphere considered here is Maxwellian, because it is not subject to extremely intense non-thermal radiation due to, for example, cosmic rays \citep[e.g.][]{ferland2016}.

Cloudy computations include a wide range of atomic (i.e. all elements up to Zn) and molecular species, and are valid across a large interval of plasma temperatures (3--10$^{10}$\,K) and densities ($<$10$^{15}$\,cm$^{-3}$), covering the parameter space of what is expected in upper planetary atmospheres. Within Cloudy, the temperature of the gas corresponds to the electron temperature.

To enable computing radiative transfer accounting for NLTE effects, Cloudy does not employ pre-computed opacity tables, because in NLTE the opacity is itself a function of the radiation field and therefore the opacities must be recomputed every time the atmospheric temperature changes from one iteration to another. Instead, Cloudy computes the opacities line-by-line, namely, at each given frequency point, the code extracts the line opacity from all available lines and adds the continuum contribution to it to obtain the total opacity. The opacity is then computed in this way at each frequency point in the considered wavelength range and for a large number of frequencies ($>$10$^6$). The frequency spacing is nearly exponentially increasing with increasing wavelength (see Cloudy user manual \#2 for more details). The spectral resolution for the Cloudy computations is given by the user and the minimum is 33. In this work, we considered a resolution of 100\,000. 

The code predicts the thermal, ionisation, and chemical structure of a gas cloud, as well as its emitted and transmitted spectrum. Cloudy predicts these quantities from specifying just (a) the shape and intensity of the external radiation field illuminating a cloud (i.e. in our case the stellar spectral energy distribution), (b) the chemical composition of the gas (we assume solar composition), and (c) the geometry of the gas, including its radial extent and the dependence of the gas density as a function of the distance to the radiation source. Cloudy does this by simultaneously solving the equations of statistical and thermal equilibrium, the equations that balance ionisation-neutralisation processes, and heating-cooling processes \citep{osterbrock2006}. Cloudy treats all ionisation stages and includes as recombination mechanisms charge-exchange, radiative recombination, and dielectronic recombination processes. The ionisation mechanisms accounted in the code comprise photoionisation from valence, inner shells, and excited states, as well as collisional ionisation by both thermal and supra-thermal electrons \citep{voronov1997,dere2007}, and charge transfer. Cloudy accounts also for molecular photodissociation.

Cloudy calculations are one-dimensional and assume plane-parallel geometry and hydrostatic equilibrium. For the specific case of KELT-9b, the latter assumption is valid within the main line formation region of optical and infrared spectral lines. This is because deviations from hydrostatic equilibrium become important once the outflow velocity reaches a significant fraction of the sound speed, that is near the sonic point, which lies close to the top boundary of the range of pressures considered here (10$^{-11}$--10$^{-12}$\,bar; see below), and thus well above the line formation region of optical and infrared spectral lines \citep[e.g.][]{koskinen2013b,turner2020,fossati2020}. Cloudy calculations are iterative and the iterations are controlled by convergence criteria on the local pressure, on the temperature (i.e. heating-cooling balance), and on the electron density, for each considered layer.

Cloudy builds up its calculations on a large database comprising model atoms (e.g. energy levels, level state definition, statistical weight of each energy level, Einstein coefficients, line oscillator strengths, collision rates, etc.) for all elements up to Zn. The information included in the model atoms is taken from \citet[][largely based on NIST\footnote{\tt https://www.nist.gov/} for the atomic part]{lykins2015}, the CHIANTI database \citep{dere1997,landi2012}, and the Leiden Atomic and Molecular Database \citep[LAMDA; ][]{schoier2005}. When computing spectral lines, Cloudy accounts for natural, thermal, and Stark broadening. Therefore, broadening through collisions with neutrals (i.e. van der Waals broadening) is not included in the code. Furthermore, Cloudy can also account for line broadening by turbulence, which modifies line opacities and the resulting optical depths, and adds a component of ram pressure to the total pressure. However, in this work, we do not include any turbulence in the calculations.
\subsubsection{Cloudy for Exoplanets}
Cloudy is a general-purpose code adaptable to a great variety of gas clouds, therefore the user has to first set up the geometry of the cloud and its basic physical conditions, namely the hydrogen density profile as a function of the radial distance to the external light source and the basic chemical composition (i.e. which elements to consider in addition to hydrogen, their abundance, and which molecules to take into account). To this end, we developed a {\sc python} interface, called ``Cloudy for Exoplanets'' (CfE), enabling one to write Cloudy input files on the basis of input parameters given by the user, run Cloudy, and read Cloudy output files, using the information contained there to set up a new Cloudy calculation in an iterative procedure until the temperature profile has converged, as described below. The details of how CfE sets up Cloudy input files and the iteration procedure are described here below.

In the simplest case, a Cloudy simulation requires as input the total hydrogen density and physical extent of the atmosphere being simulated, and a source of illumination. In addition to this, for computing the temperature and chemical structure of an exoplanetary atmosphere, CfE requires the planetary and stellar masses ($M_p$ and $M_\star$) and radii ($R_p$ and $R_\star$), the orbital semi-major axis ($a$), the orbital period ($\tau$), an approximate planetary equilibrium temperature ($T_\mathrm{eq}$), the optical continuum pressure level ($p_0$), and the bulk atmospheric metallicity ($[M/H]$). With this information, CfE creates an initial sub-stellar one-dimensional (1D) atmospheric structure by assuming an isothermal atmosphere of $T$\,=\,$T_\mathrm{eq}$, scaled solar abundance ratios, no ionisation or excitation, and computes the radius scale over a given pressure range assuming hydrostatic equilibrium as
\begin{equation}
\label{eq.radius_H_P}
     r_{i+1} = -H_i\ln{\left(\frac{P_{i+1}}{P_i}\right)} + r_i\,,
\end{equation}
where $r_i$ is the radius from the planet's center of the $i$th level, $H_i$ is the pressure scale height of the $i$th level, and $P_i$ is the atmospheric pressure of the $i$th level. The pressure scale height is calculated as
\begin{equation}
     H_i = \frac{kT_{i+\frac{1}{2}}}{m_{i+\frac{1}{2}}g_i}\,,
\end{equation}
where $k$ is Boltzmann's constant, $T_{i+\frac{1}{2}}$ is the average atmospheric temperature between the $i$th and $i$th$+1$ levels, $m_{i+\frac{1}{2}}$ is the average mean molecular weight of the atmosphere between the $i$th and $i$th$+1$ levels, and $g_i$ is the gravitational acceleration experienced by atmospheric particles at the $i$th level. 

To solve Equation~\ref{eq.radius_H_P}, one requires to adopt an appropriate reference $p_0$--$R_p$ pair to set the value of the planetary radius corresponding to the optical continuum pressure level ($p_0$). There is a well known pressure-radius reference location degeneracy in this choice \citep[e.g.][]{lecavelier2008,benneke2012,dewit2013,griffith2014,heng2017,betremieux2017}. We note that \citet{heng2017} extends this degeneracy into a three-way degeneracy by including the water abundance. However, \citet{welbanks2019} later showed that the degeneracy between planetary radius and its reference pressure is well characterised irrespective of the water abundance, and has little effect on abundance estimates. Therefore, the $p_0$--$R_p$ choice, assuming it is reasonably close to reality, does not have a significant impact on the results (see below).

The gravitational acceleration accounts for the effects of the Roche potential, formulated as
\begin{multline}
     g_{r,i} = -\frac{GM_p}{r_i^2} -\frac{GM_\star(r_i \lambda^2-a\lambda+r_i*\xi^2+r_i*\nu^2)}{[(a - r_i \lambda)^2 + (r_i \xi)^2 + (r_i \nu)^2]^\frac{3}{2}} + \\ \omega^2(r_i\lambda^2 - \mu a \lambda + r_i\xi^2)\,,
\end{multline}
\begin{multline}
     g_{\theta,i} = -\frac{GM_\star(\lambda_p (r_i
\lambda-a)+r_i\xi\xi_p+r_i\nu\nu_p)}{[(a - r_i \lambda)^2 + (r_i \xi)^2 + (r_i \nu)^2]^\frac{3}{2}} + \omega^2(\xi_p(r_i\lambda - \mu a)+ r_i\xi\xi_p)\,,
\end{multline}
\begin{equation}
     g_{\phi,i} = \frac{GM_\star a\xi}{\nu_p[(a - r_i \lambda)^2 + (r_i
\xi)^2 + (r_i \nu)^2]^\frac{3}{2}} -\frac{\omega^2\mu a\xi}{\nu_p}\,,
\end{equation}
and
\begin{equation}
     g_i = \sqrt{g_{r,i}^2+g_{\theta,i}^2+g_{\phi,i}^2}\,,
\end{equation}
where $G$ is the gravitational constant, $\omega$\,=\,$2\pi/\tau$ is the orbital angular frequency, $\mu$\,=\,$\frac{M_\star}{M_\star+M_p}$, and $\lambda$\,=\,$\cos{(\phi)}\sin{(\theta)}$, $\lambda_p$\,=\,$\cos{(\phi)}\cos{(\theta)}$, $\xi$\,=\,$\sin{(\phi)}\sin{(\theta)}$, $\xi_p$\,=\,$\cos{(\phi)}\cos{(\theta)}$,
$\nu$\,=\,$\cos{(\theta)}$, $\nu_p$\,=\,$-\sin{(\theta)}$, for co-latitude $0\leq\theta\leq\pi$ and longitude $0\leq\phi\leq 2\pi$. This simplifies to
\begin{equation}
      g_i = -\frac{GM_p}{r_i^2} +\frac{GM_\star}{(a-r_i)^2} + \omega^2(r_i - a\frac{M_\star}{M_\star+M_p})
\end{equation}
for $\theta$\,=\,$\pi/2$ and $\phi$\,=\,0 at the substellar point.

This initial radius scale and density profile are passed to Cloudy, which computes the thermal and chemical equilibrium solution of the atmosphere, and the resulting output is used to generate a new atmospheric structure following the above method, but without assuming an isothermal and neutral atmosphere. In practice, CfE updates the hydrogen density structure of the planetary atmosphere given as input to Cloudy employing the temperature and mean molecular weight as a function of pressure obtained from the previous Cloudy run. The process is repeated for a number of iterations set by the user, after which the converged temperature structure, chemical solution, and radius scale of the atmosphere are recorded, as well as the radius scale at the terminator, computed from the final structure according to the above method using polar latitude and longitude coordinates. Experience indicates that the temperature structure stabilises following two CfE iterations, but we run CfE up to four iterations reaching a $<$5\,K difference between the TP profiles computed in the last two CfE runs.

The Cloudy atmospheric model indicates that in the case of KELT-9b the maximum gas density above which Cloudy results become unreliable lies around the one mbar level. Therefore, for the CfE calculations, we set $p_0$ equal to 0.175\,bar, that is the pressure level at which the HELIOS model gives an optical depth of 2/3 integrating over the KELT bandpass filter, which barycenter lies at $\approx$6000\,\AA\ \citep{pepper2007}. Furthermore, to solve Equation~\ref{eq.radius_H_P} and to be consistent with the estimate of $p_0$, we adopted r($p_0$) equal to the observed transit radius $R_p$\,=\,1.936\,$R_{\rm Jup}$ given by \citet{borsa2019}, which is based on the transit depth measurement obtained from the KELT light curves \citep{gaudi2017}. We tested the impact of this choice on the obtained TP profile by recomputing the Cloudy TP profile employing smaller radii of 1.6, 1.65, 1.7, 1.75, 1.8, and 1.85\,$R_{\rm Jup}$ (instead of 1.936\,$R_{\rm Jup}$), obtaining always an identical result (Figure~\ref{fig:TP_Rp}).

For all Cloudy calculations, we took into account all elements up to Zn and only hydrogen molecules (i.e. H$_2$, H$^{+}_{2}$, and H$^{+}_{3}$), because other molecules play a minor role in the hot environment of KELT-9b's atmosphere in determining the atmospheric physical properties \citep[e.g.][]{lothringer2019}.

Cloudy computations do not enable one to consider atmospheric heat redistribution, therefore we mimicked $f$ by adding a scaling factor to the input stellar spectral energy distribution that we varied until the Cloudy and HELIOS TP profiles matched around the 10$^{-4}$\,bar level. This is close to the highest pressure at which Cloudy calculations are still valid and NLTE effects are small \citep{fossati2020}, making Cloudy and HELIOS results comparable. Indeed, in this part of the atmosphere, the two TP profiles follow the same shape. We computed TP profiles employing a range of scaling factors and obtaining that the one leading to best match the HELIOS profile around the 10$^{-4}$\,bar level is 0.48. We also compared the TP profiles computed with different values of the scaling factor obtaining that the TP profile at pressures below 10$^{-5}$\,bar, thus in the main line formation region, is only weakly dependent on it and increases by about 200\,K with the scaling factor going from 0.4 to 1. Therefore, we conclude that the general results presented here can be considered to be robust against the choice of the scaling factor. The adopted TP profile obtained with CfE, and accounting for NLTE effects, is shown in Figure~\ref{fig:compositeTP}. 
\section{Results}\label{sec:results}
We defined the final TP profile over the 10--10$^{-11}$\,bar pressure range, dividing it into 180 layers equally spaced in $\log{p}$. This pressure range is wide enough to fully contain the atmospheric formation region of ultraviolet to infrared spectral lines. It further ensures that the atmosphere is transparent to light at ultraviolet-to-infrared wavelengths at the top of the atmosphere and opaque in the same wavelength range at the bottom of the atmosphere. We run models considering a smaller and larger number of atmospheric layers, but the results are identical to those obtained considering 180 layers (see Figure~\ref{fig:TPlayers}).

We joined the HELIOS and Cloudy TP profiles in the 1.4$\times$10$^{-4}$ and 6.0$\times$10$^{-5}$\,bar pressure range, which is where the shape of the two profiles looks most similar. In this pressure range, we derived the composite TP profile by averaging between the HELIOS and Cloudy results, further interpolating on the final pressure scale. At pressures higher than 1.4$\times$10$^{-4}$\,bar and lower than 6.0$\times$10$^{-5}$\,bar, we set the composite TP profile to correspond to those obtained with HELIOS and Cloudy, respectively, also interpolating them on the final pressure scale. Figure~\ref{fig:compositeTP} shows the composite TP profile, in comparison to the HELIOS, Cloudy, and PHOENIX TP profiles.

By construction, at high pressures the composite TP profile is in agreement with the PHOENIX TP profile and with the measured day-side temperature. The profile further presents a rather steep temperature increase between about 1 and 10$^{-7}$\,bar in which the temperature goes from $\approx$4200\,K to $\approx$8400\,K, and it remains roughly constant at lower pressures. In the upper atmosphere, the temperature is about 1800--2500\,K higher than what predicted by {\sc HELIOS} and {\sc PHOENIX}. The shape of the composite TP profile goes in the direction of the empirical TP profile \citet{fossati2020} found to best match the observed hydrogen Balmer lines. In particular, in the lower atmosphere the TP profile empirically obtained by \citet{fossati2020} is about 1000--2000\,K hotter than what obtained here, but that profile was unconstrained by the observations outside of the main line formation region, namely at pressures higher and lower than roughly 10$^{-3}$ and 10$^{-8}$\,bar, respectively (Figure~\ref{fig:compositeTP}). As a matter of fact, both profiles concur on the presence of a rather steep temperature rise starting at the 10\,mbar pressure level. The main difference lies at the top of the upper atmosphere, where the TP profile presented here is about 2000\,K cooler than the empirical TP profile of \citet{fossati2020} and is roughly isothermal at pressures lower than 10$^{-7}$\,bar, instead of constantly increasing with decreasing pressure. Within the main line formation region the temperature changes by almost 3000\,K, which indicates that the assumption of an isothermal atmosphere, for example taken by \citet{wyttenbach2020} and \citet{yan2020a}, is most likely unphysical and should be avoided.

\begin{figure}[h!]
\includegraphics[width=\hsize,clip]{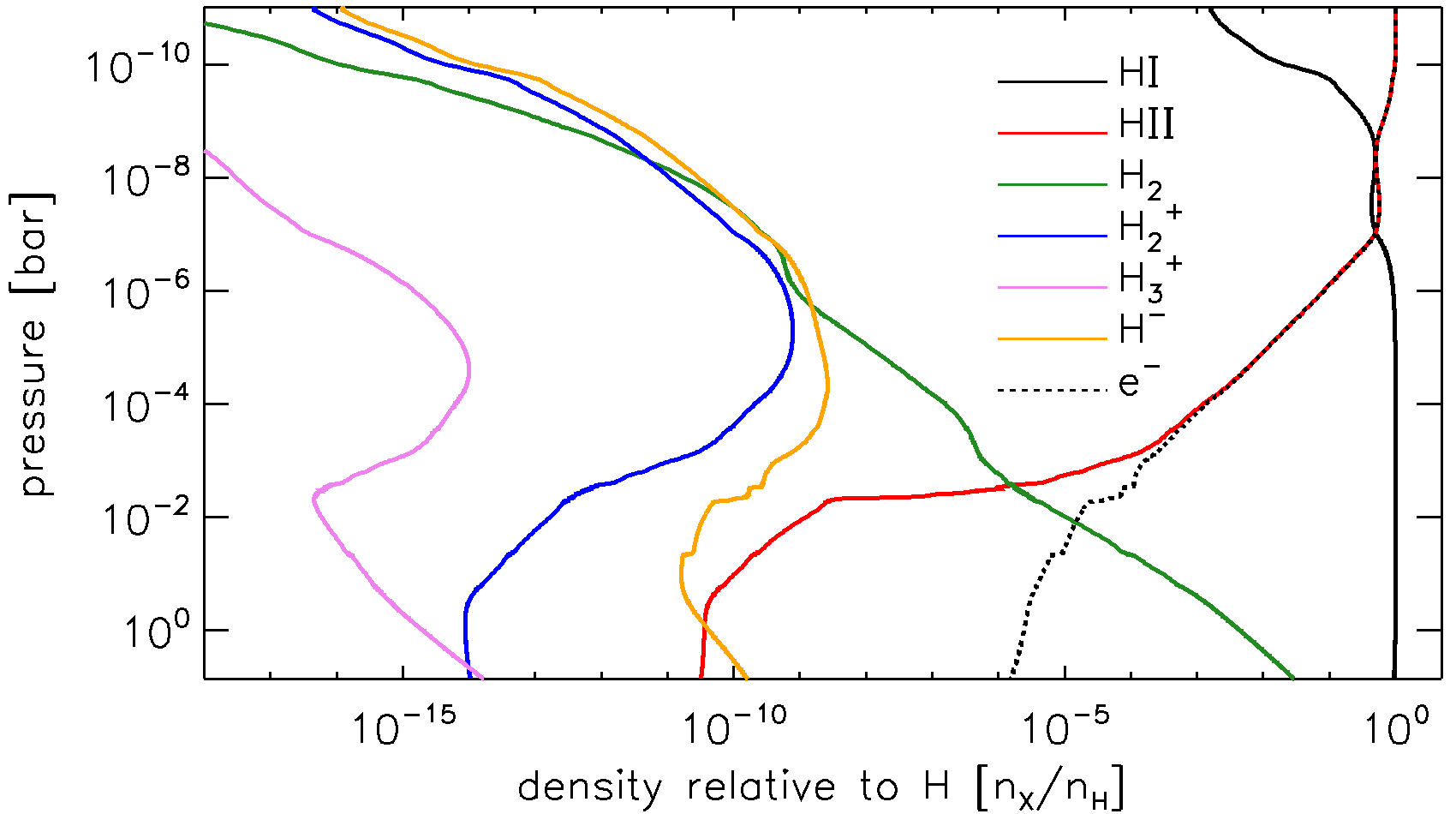}
\includegraphics[width=\hsize,clip]{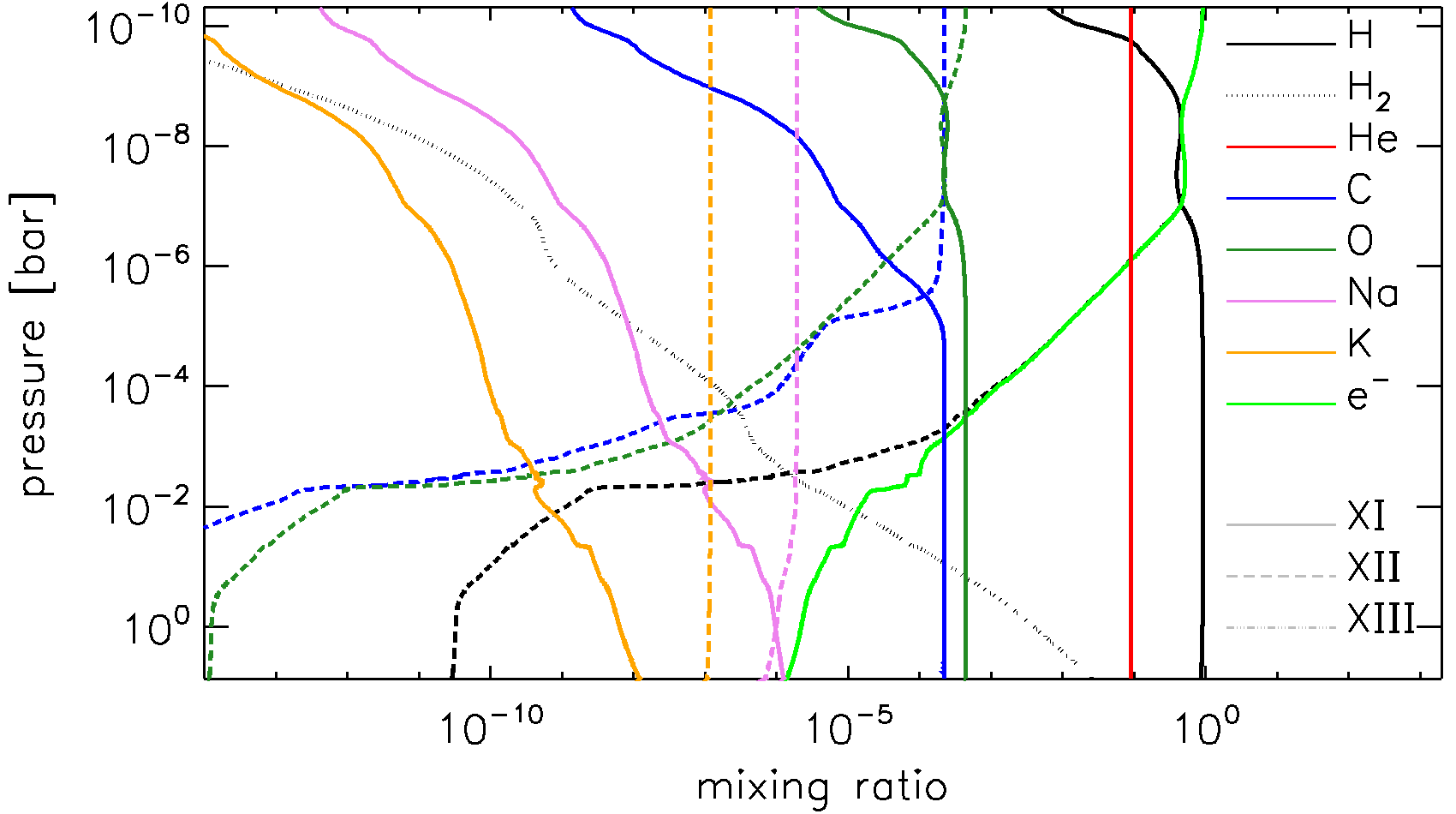}
\includegraphics[width=\hsize,clip]{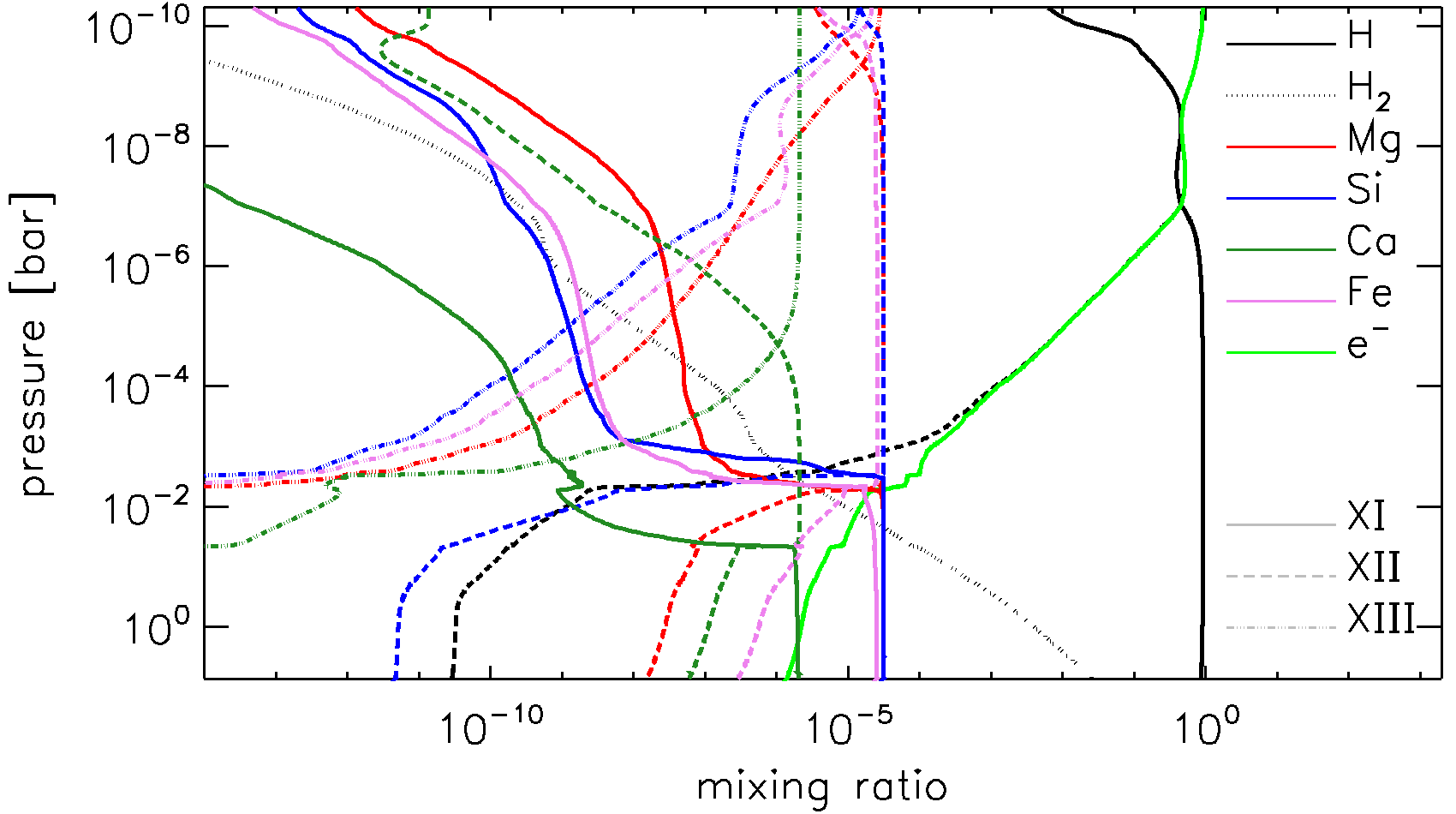}
\caption{Top: Density relative to the total density of hydrogen for neutral hydrogen (H{\sc i}; black solid), protons (H{\sc ii}; red), molecular hydrogen (H$_2$; dark green), H$_2^+$ (blue), H$_3^+$ (violet), H$^-$ (orange), and electrons (e$^-$; black dashed). Middle: Mixing ratios for hydrogen (black), H$_2$ (black-dotted), He (red), C (blue), O (dark green), Na (violet), K (orange), and electrons (bright green) as a function of atmospheric pressure. Neutral (X{\sc i}), singly ionised (X{\sc ii}), and doubly ionised (X{\sc iii}) species are marked by solid, dashed, and dash-dotted lines, respectively. Bottom: Same as middle panel, but for Mg (red), Si (blue), Ca (dark green), and Fe (violet). The hydrogen, H$_2$, and e$^-$ mixing ratios are shown in both middle and bottom panels for reference.}
\label{fig:chemistry}
\end{figure}

To also obtain a homogeneous chemical atmospheric structure, the composite TP profile is passed to CfE to generate input for a single iteration while fixing the temperature structure in the Cloudy simulation. The top panel of Figure~\ref{fig:chemistry} shows the details of the atmospheric composition for hydrogen bearing species (neutral hydrogen H{\sc i}, protons H{\sc ii}, H$^-$, molecular hydrogen H$_2$, H$_2^+$, H$_3^+$, electrons e$^-$) with respect to the total hydrogen density as a function of pressure. The atmosphere is dominated by neutral hydrogen at pressures higher than about 10$^{-7}$\,bar, while protons are the most abundant species at lower pressures. In the lower atmosphere, at pressures higher than about 10$^{-3}$\,bar, H{\sc i} and H$_2$ dominate the hydrogen atmospheric composition, with the latter decreasing rapidly with decreasing pressure. H$^-$ is the fourth most abundant hydrogen species in the lower atmosphere and dominates the spectral shape of the continuum \citep[e.g.][]{arcangeli2018}. The H{\sc ii} abundance increases quickly with decreasing pressure starting from the 10$^{-2}$\,bar level. This is mostly due to thermal ionisation, because hydrogen photoionisation occurs at higher altitudes and it is believed to be small due to the shape of the stellar spectral energy distribution that is characterised by a very weak emission of extreme ultraviolet radiation \citep{fossati2018}.

The middle and bottom panels of Figure~\ref{fig:chemistry} show instead the mixing ratio as a function of pressure for some of the species most relevant in terms of abundance and observability at ultraviolet and/or optical wavelengths. As a consequence of the solar composition assumption, He{\sc i} is the second most abundant species throughout most of the atmosphere. Carbon and Oxygen behave in a similar way remaining mostly in neutral form up to the 10$^{-5}$ and 10$^{-7}$\,bar level, respectively, being then mostly singly ionised at lower pressures. This result suggests that the several C (neutral and ionised) and O (mostly neutral) resonance lines lying in the far ultraviolet might be detectable in transmission, for example with HST. Because of their similar configuration, Na and K behave alike and are singly ionised throughout most of the atmosphere.

As a result of their similar ionisation potentials, Mg, Si, and Fe behave alike, with the neutral species dominating at pressures higher than about 5\,mbar and the singly ionised species dominating at lower pressures, up to the $\approx$10$^{-9}$\,bar level where the doubly-ionised species take over. Single spectral lines of these species, particularly Mg and Fe, have been directly detected in transmission spectra of KELT-9b \citep[e.g.][]{cauley2019}. The near-ultraviolet wavelength range contains several strong Mg, Si, and Fe lines that could be detectable with both HST and CUTE \citep{fleming2018}. The distribution of Ca atoms presents a clear separation, with Ca{\sc i} dominating at pressures higher than about 0.1\,bar, Ca{\sc ii} dominating up to the 10$^{-5}$\,bar level, and Ca{\sc iii} dominating at lower pressures. However, the Ca{\sc ii} density decreases rather slowly with decreasing pressure above the 10$^{-5}$\,bar level, namely in the main line formation region, probably explaining why the Ca{\sc ii} infrared triplet has been detected in transmission \citep{turner2020}.

The density profiles shown in Figure~\ref{fig:chemistry} are remarkably similar to those obtained by \citet{fossati2020} and extracted with Cloudy on the basis of one of the empirical TP profiles they found to better fit the H$\alpha$ and H$\beta$ lines. The main difference lies in the slightly higher pressure levels at which ionised species become more abundant than neutral species. This is driven by the cooler temperature obtained here, through forward modelling, compared to what obtained by \citet{fossati2020}, though a grid approach.
\section{Discussion}\label{sec:discussion}
\subsection{Origin of the upper atmospheric heating}\label{sec:nlte}
\subsubsection{NLTE vs LTE temperature-pressure profile}\label{sec:lte_TP}
\citet{fossati2020} found that the family of TP profiles best fitting the hydrogen Balmer lines is characterised by an inverted temperature profile with an upper atmospheric temperature of the order of 10000\,K. The high ionisation fraction of metals in the upper atmosphere found on the basis of the TP profiles best fitting the hydrogen Balmer lines, led them to propose metal photoionisation as the upper atmospheric heating mechanism. However, thanks to the implementations described in Section~\ref{sec:lower}, the HELIOS TP profile has been computed accounting for metal photoionisation. Therefore, the origin of the extra heating lies somewhere else.

One of the aspects separating HELIOS from Cloudy is the assumption of LTE opacity in the HELIOS model calculations. We remark that also the PHOENIX TP profile has been computed assuming LTE. Therefore, under the assumption of radiative equilibrium, NLTE effects may be a viable cause for the high upper atmospheric temperature. In particular, the overpopulation and/or underpopulation of specific energy levels of species playing a major role in the atmospheric opacity, and thus heating and cooling, may significantly affect the shape of the TP profile. Therefore, we made use of Cloudy, which has the capability of running computations in both LTE and NLTE, to identify the impact of NLTE effects on the TP profile. We remark that the LTE TP profile still accounts for photoionisation.

\begin{figure}[h!]
\includegraphics[width=\hsize,clip]{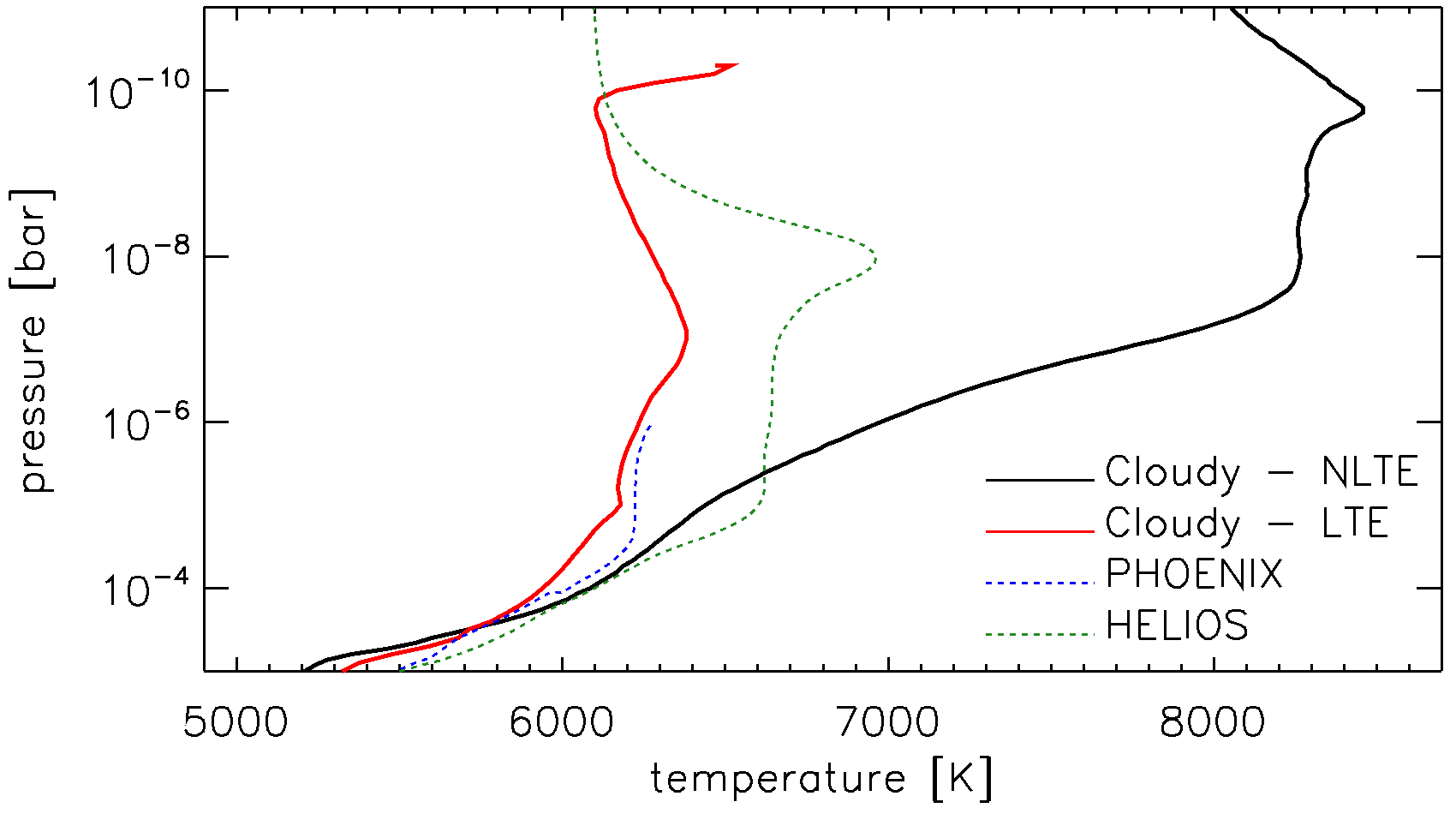}
\caption{Temperature-pressure profiles of the upper atmosphere computed with the Cloudy code accounting for NLTE (black; same as in Figure~\ref{fig:compositeTP}) and assuming LTE (red). The PHOENIX (blue dashed line) and HELIOS (green dashed line) TP profiles are also shown for reference.}
\label{fig:lteVSnlte}
\end{figure}

Figure~\ref{fig:lteVSnlte} shows the TP profiles computed with Cloudy in LTE and NLTE, also in comparison with those obtained with PHOENIX and HELIOS assuming LTE. At the top of the atmosphere, the LTE TP profile ends at somewhat higher pressure compared to the NLTE TP profile, because Cloudy experienced convergence problems at very low gas densities and assuming LTE. In general, the shape of the LTE Cloudy TP profile resembles that of the PHOENIX and HELIOS TP profiles, but it is slightly cooler and does not show the rather narrow temperature bump around 10$^{-8}$\,bar that characterises the HELIOS TP profile.

There is a significant difference between the LTE and NLTE Cloudy TP profiles. At pressures higher than about 10$^{-4}$\,bar the two profiles are similar, but diverge significantly at lower pressures. In the upper atmosphere, thus in the main formation region of major spectral lines \citep{turner2020,fossati2020}, the TP profile computed accounting for NLTE effects is more than 1000\,K hotter than the LTE one, with the largest difference reaching more than 2000\,K at pressures ranging between about 5$\times$10$^{-8}$ and 10$^{-10}$\,bar.
\subsubsection{Atmospheric heating and cooling processes}\label{sec:processes}
\begin{figure}[ht!]
\includegraphics[width=\hsize,clip]{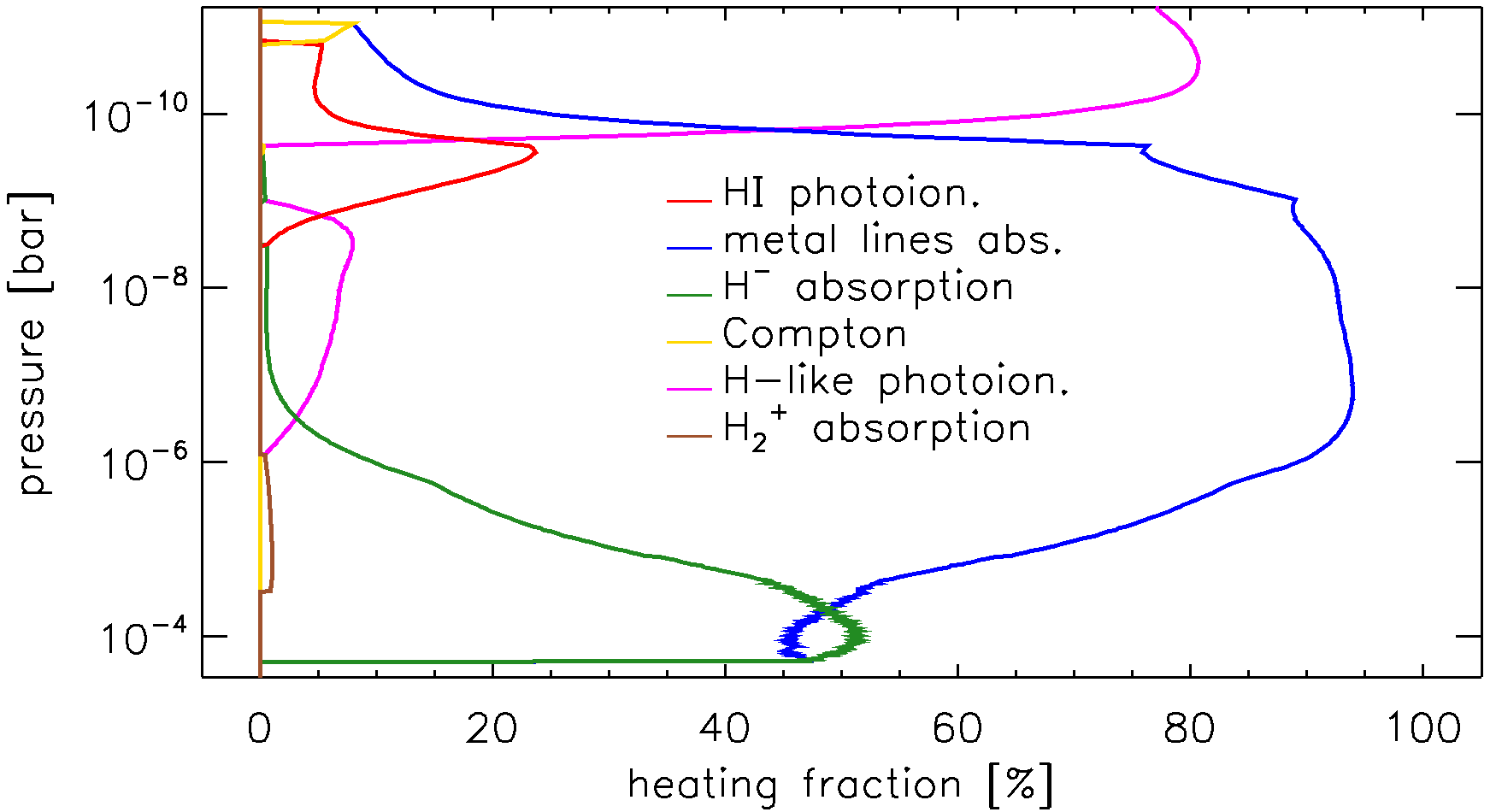}
\includegraphics[width=\hsize,clip]{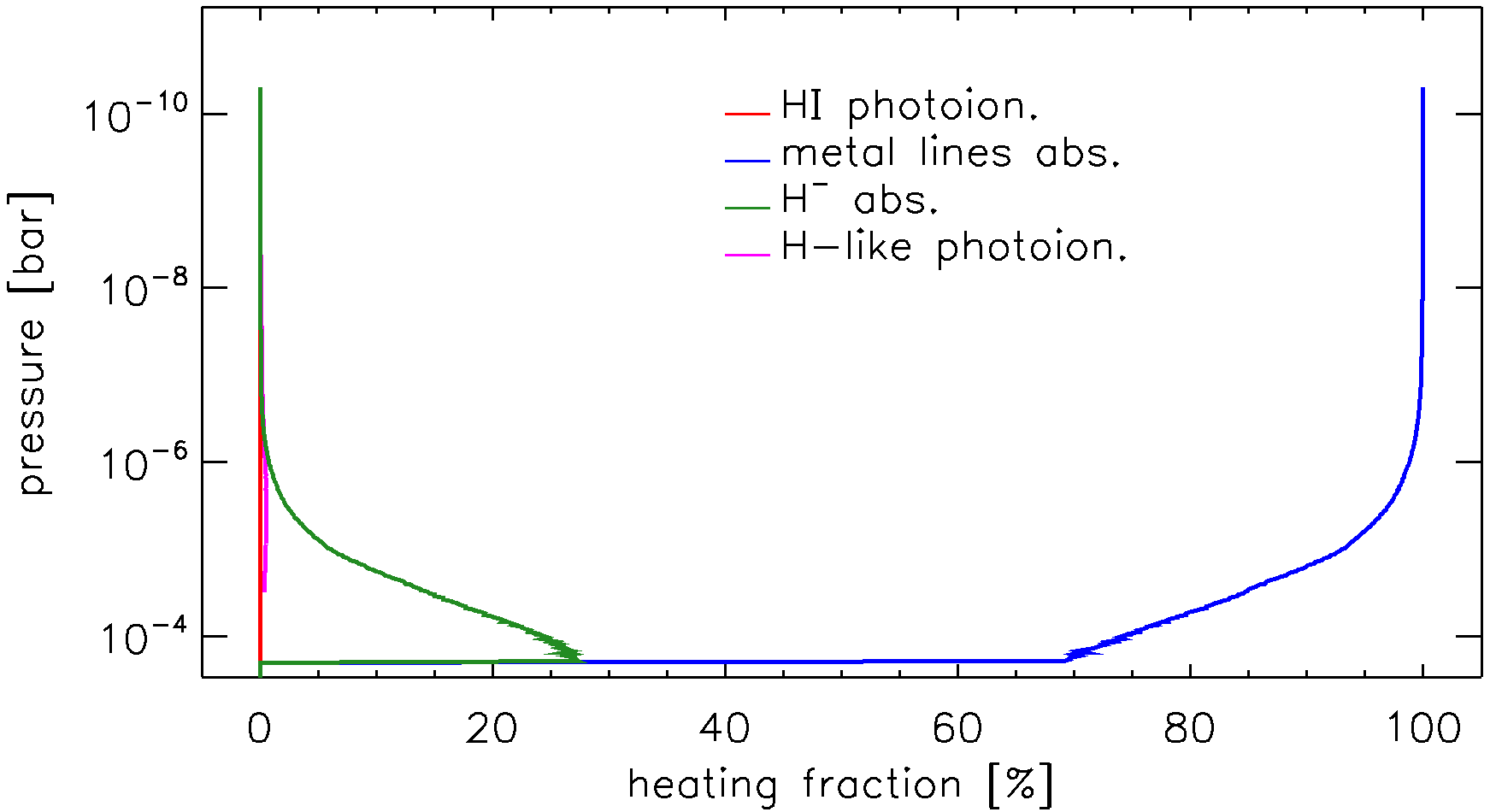}
\caption{Top: heating contribution (in \%) of the three most important heating processes at each atmospheric layer as a function of pressure extracted from the NLTE Cloudy run. The relevant heating processes occurring in the upper atmosphere are hydrogen photoionisation (red; photoionisation of H{\sc i} lying in the ground state), metal line absorption (blue), H$^-$ absorption (green), Compton heating (i.e. electron absorption; yellow), photoionisation of hydrogenic species (magenta; photoionisation of excited H{\sc i}), and H$_2^+$ absorption (brown). Metal absorption is the main heating contributor at pressures ranging between about 5$\times$10$^{-5}$ and 10$^{-10}$\,bar, where the atmospheric temperature increases steeply and reaches its maximum. The heating fraction of any given process is artificially set equal to zero in the atmospheric regions in which the heating process is not among the three most important ones. Bottom: same as top, but for the LTE Cloudy run.}
\label{fig:heating}
\end{figure}

\begin{figure}[h!]
\includegraphics[width=\hsize,clip]{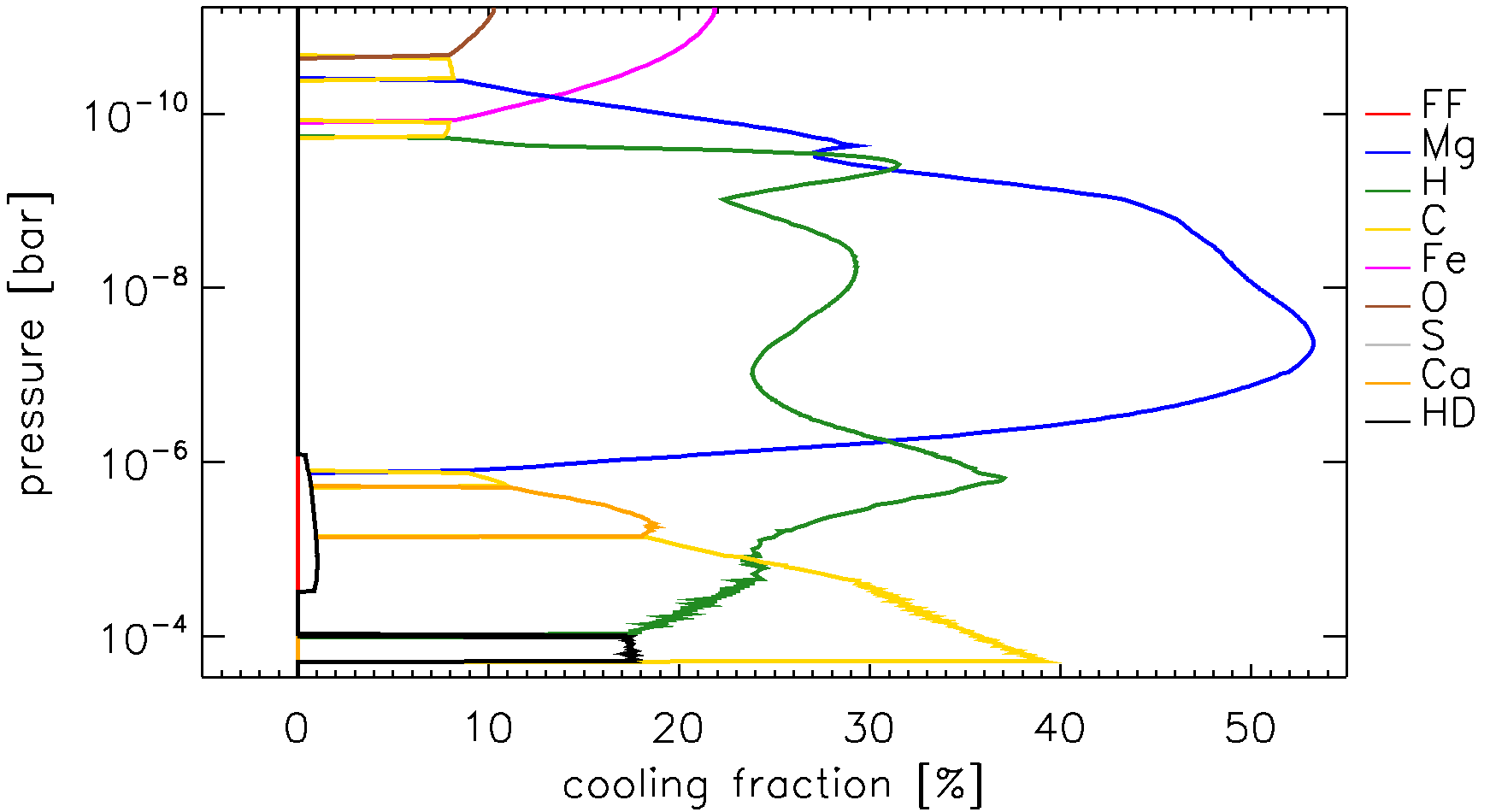}
\includegraphics[width=\hsize,clip]{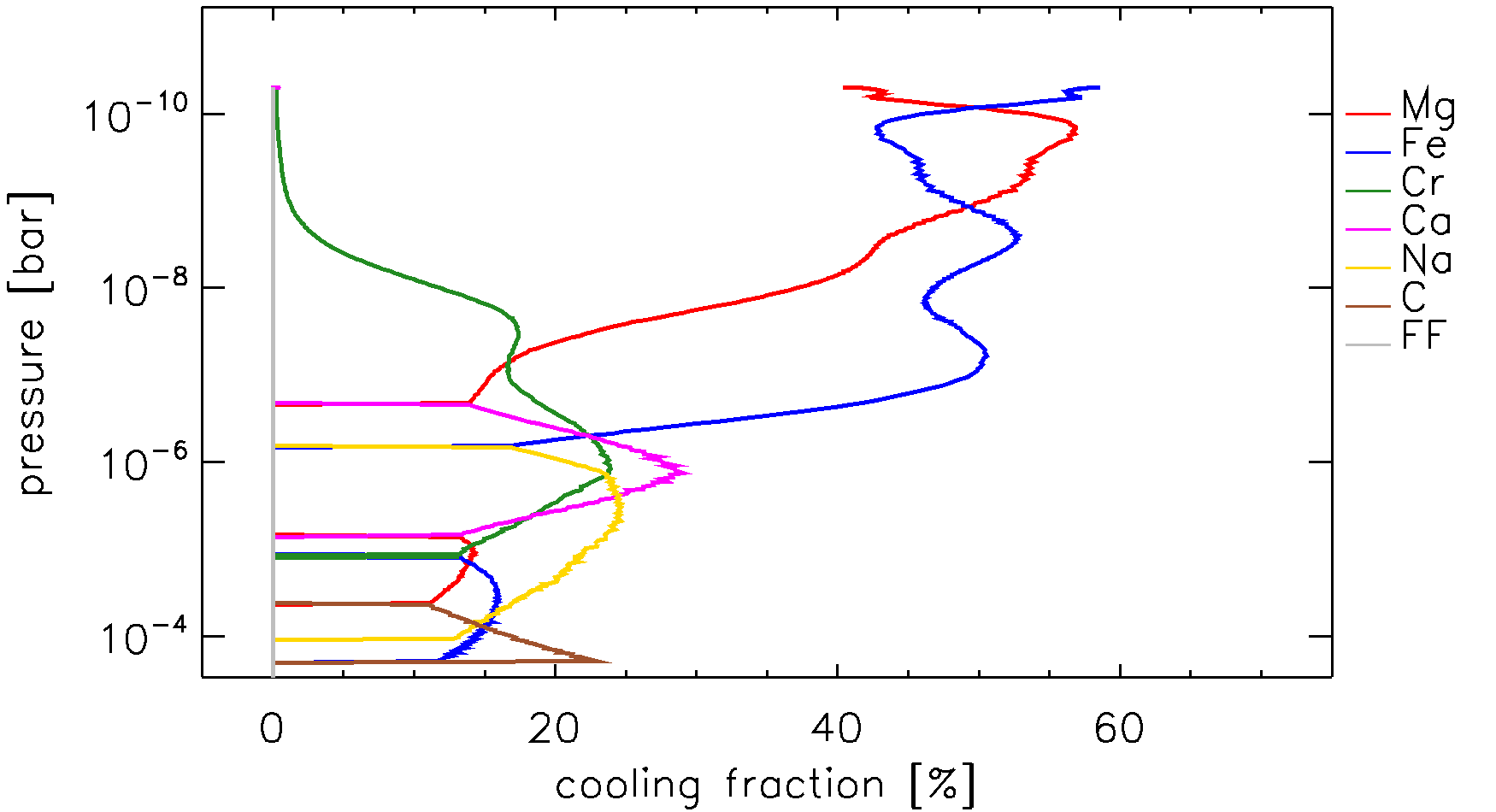}
\caption{Cooling contribution (in \%) of the three most important cooling processes or species at each atmospheric layer as a function of pressure obtained from the Cloudy NLTE (top) and LTE (bottom) runs. The label ``FF'' indicates free-free cooling from H and He. The cooling fraction of any given species is artificially set equal to zero in the atmospheric regions in which the species producing the cooling is not among the three most important ones. Magnesium is the species most contributing to the cooling of the upper atmosphere.}
\label{fig:cooling}
\end{figure}

We explored the Cloudy models to identify the species and energy levels responsible for the extra heating and/or lack of cooling driving the difference between the LTE and NLTE TP profiles. Figures~\ref{fig:heating} and \ref{fig:cooling} show respectively the heating and cooling contribution provided by the three most important heating and cooling processes in the atmosphere as a function of pressure extracted from the Cloudy output obtained from the NLTE and LTE runs. In both NLTE and LTE cases, metal line absorption is the largest heating contribution at pressures lower than $\sim$10$^{-4}$\,bar, except for the very top of the atmosphere in the NLTE case where the photoionisation of hydrogenic species (i.e. photoionisation of excited H{\sc i}, in this case) is the major heating contribution, though metal line absorption is still significant. Absorption from H$^-$ becomes progressively more important as pressure increases, but also at these deeper layers metal line absorption heating is significant, and it is still the main heating contributor in the LTE case.

\begin{figure}[h!]
\includegraphics[width=\hsize,clip]{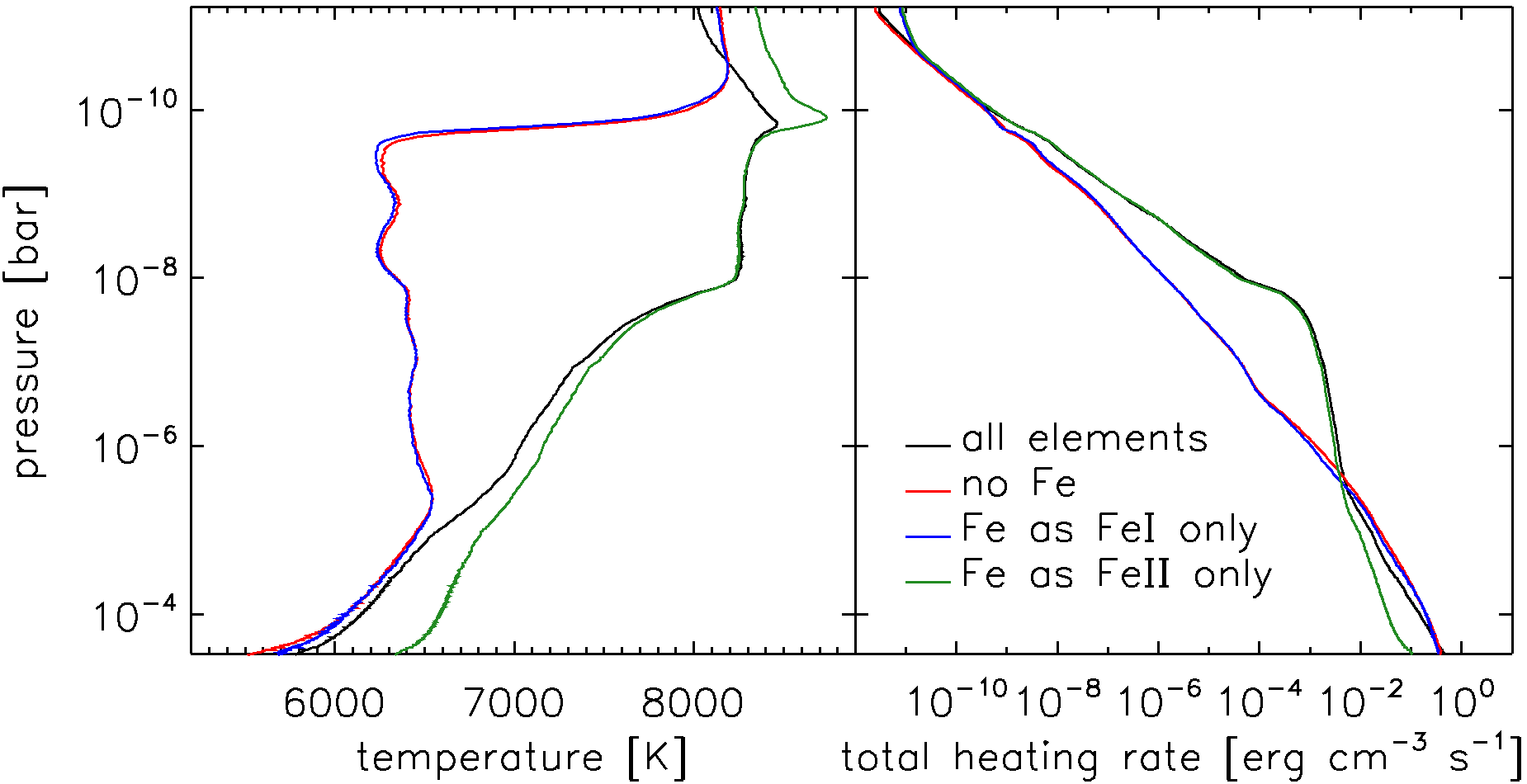}
\caption{Temperature (left) and total heating rate (right) as a function of pressure obtained accounting for all elements (black; this is the same TP profile shown in Figures~\ref{fig:compositeTP} and \ref{fig:lteVSnlte}), considering Fe only in the form of Fe{\sc i} (blue; i.e. Fe{\sc i} is not allowed to ionise), considering Fe only in the form of Fe{\sc ii} (green; i.e. Fe{\sc ii} is not allowed to recombine or ionise), and after removing Fe among the list of considered elements (red).}
\label{fig:Feheating}
\end{figure}

Although Cloudy does not give information on the total heating rate due to each species, we inferred this for Fe in NLTE by taking the output obtained after the last CfE iteration and then running a further iteration considering only Fe{\sc i} (i.e. Fe{\sc i} is not allowed to ionise), or considering only Fe{\sc ii} (i.e. Fe{\sc ii} is not allowed to ionise or recombine), or removing Fe completely from the list of considered atmospheric species. For the cases in which we fixed Fe corresponding to Fe{\sc i}/Fe{\sc ii}, we employed the Fe{\sc i}/Fe{\sc ii} density profile obtained from the model that accounts for all elements. Figure~\ref{fig:Feheating} shows the temperature and the total heating rate as a function of pressure obtained in these four cases. By removing Fe, the heating rate decreases significantly in the 10$^{-5}$ to 10$^{-10}$\,bar pressure range, that is precisely the portion of atmosphere most differing between the LTE and NLTE case. Figure~\ref{fig:Feheating} clearly indicates that most of the heating in this part of the atmosphere is caused by Fe{\sc ii}.

The heating rate around the nbar level and above is comparable to what is typically obtained from hydrodynamic modeling of classical hot Jupiters \citep[e.g.][]{yelle2004,koskinen2013a}. However, at higher pressures, particularly when including Fe, the heating rate is much higher. This is due to the metals included in Cloudy that are mostly missing from current hydrodynamic upper atmosphere models. This result demonstrates that, if Fe reaches the upper atmosphere, accounting for Fe and NLTE effects becomes of critical importance for correctly modelling the atmospheric energy balance. This is particularly relevant for planets orbiting stars with a spectral energy distribution that peaks in the FUV/NUV range, which hosts most of the strong Fe{\sc ii} lines originating from low energy levels.

Figure~\ref{fig:cooling} shows that numerous processes and species concur in cooling the upper atmosphere both in NLTE and LTE. However, in general, Figure~\ref{fig:cooling} indicates that Mg is the most important atmospheric coolant in both cases.
\subsubsection{The role of Mg and Fe in the TP profile}\label{sec:species}
\begin{figure}[h!]
\includegraphics[width=\hsize,clip]{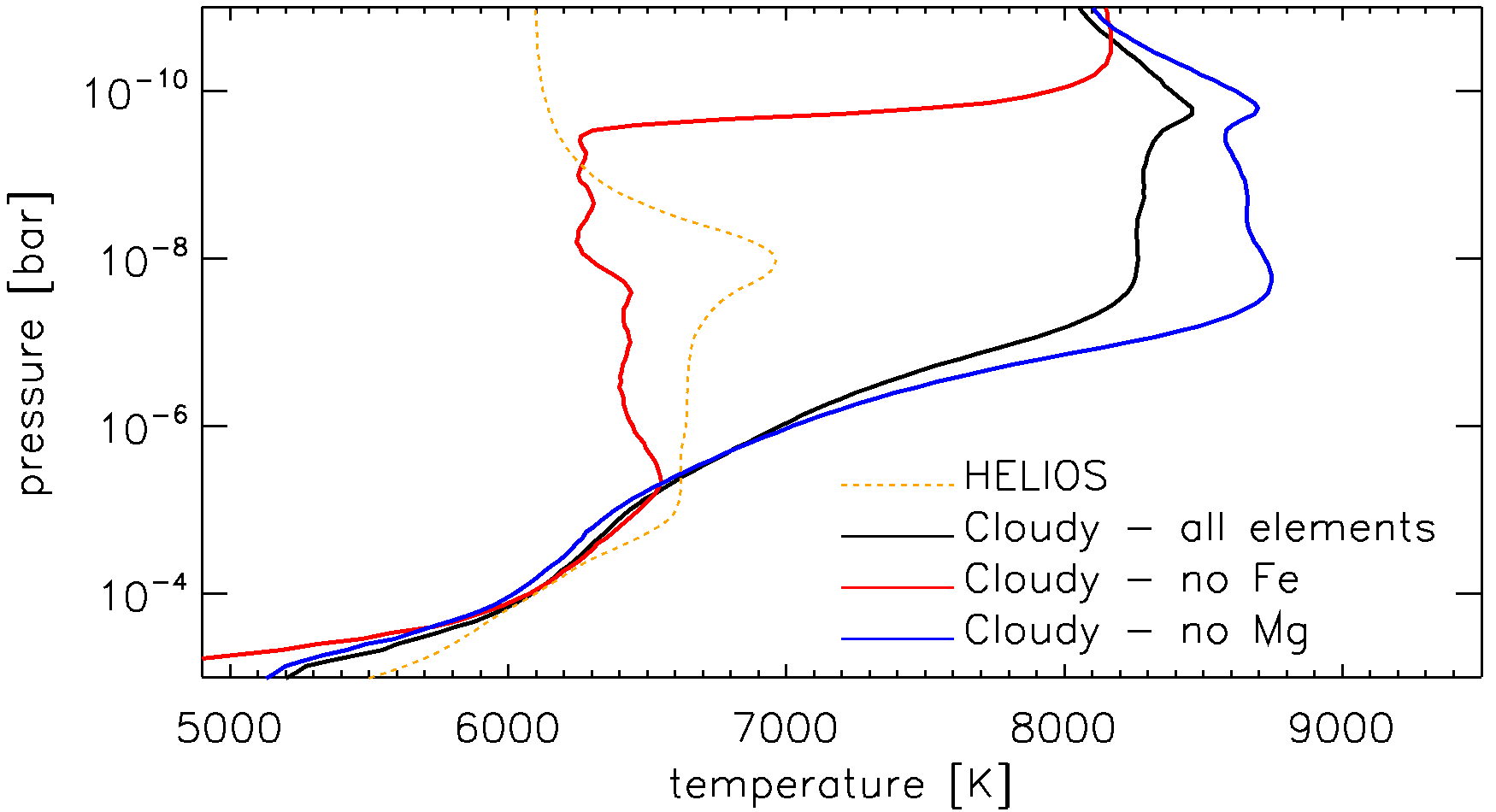}
\caption{Comparison between the NLTE Cloudy TP profiles obtained considering all elements up to Zn (black solid line) and excluding Fe (red solid line) or Mg (blue solid line). The orange dashed line shows the HELIOS (i.e. LTE) TP profile for reference.}
\label{fig:TP_noFeMg}
\end{figure}

To quantify the role of Fe and Mg in heating and cooling the planetary atmosphere, we computed TP profiles with CfE in NLTE excluding either Fe or Mg from the list of considered elements. This is different from what shown in Figure~\ref{fig:Feheating}, because in that case the calculation was based on just one Cloudy iteration for which the starting point was the TP profile computed accounting for all elements, while here it is the full CfE run that does not consider Fe among the list of elements. This is to ensure the maximum possible consistency when comparing results. Figure~\ref{fig:TP_noFeMg} compares the Cloudy TP profiles obtained considering all elements up to Zn with those computed excluding Fe or Mg.

Excluding Fe from the Cloudy calculation has led to an almost 2000\,K cooler TP profile at pressures ranging between 10$^{-5}$ and 5$\times$10$^{-10}$\,bar, which Figure~\ref{fig:heating} shows being heated predominantly by metal line absorption. This indicates that there must be numerous Fe lines, likely rising from overpopulated energy levels, contributing to the heating in this part of the atmosphere. Even in absence of Fe, the temperature rises steeply at the top of the atmosphere, which is due to the increasing importance of H{\sc i} photoionisation heating with decreasing pressure (Figure~\ref{fig:heating}, top-left panel). 

Excluding Mg from the computation of the TP profile has the opposite effect of excluding Fe, though to a smaller extent. The largest difference between the TP profiles computed with and without Mg is found in the 10$^{-7}$ to 10$^{-10}$\,bar pressure range, with the maximum difference of about 500\,K being at a pressure of about 5$\times$10$^{-8}$\,bar. As expected, this is the location at which the top panel of Figure~\ref{fig:cooling} indicates that Mg is the most important cooling species. 

We performed similar calculations excluding each of the other elements one at a time obtaining differences that are significantly smaller than those obtained by removing Mg or Fe, and in most cases the differences are negligible. Furthermore, the Mg and Fe abundance profiles extracted from the LTE and NLTE models are similar. Therefore, the overpopulation of certain levels of Fe and/or the underpopulation of certain levels of Mg is likely to be responsible for the large difference between the NLTE and LTE TP profiles.

This is confirmed by the departure coefficients ($b$) we extracted from the Cloudy NLTE run for Fe{\sc i}, Fe{\sc ii}, Fe{\sc iii}, Mg{\sc i}, and Mg{\sc ii}. The departure coefficients are defined as
\begin{equation}
b\,=\,\frac{n_{\rm NLTE}}{n_{\rm LTE}}\,,
\end{equation}
where $n_{\rm NLTE}$ and $n_{\rm LTE}$ are the densities of a given atom lying in a certain level in NLTE and LTE, respectively. The n$_{\rm LTE}$ profiles are those obtained through the Boltzmann equation.

Figures~\ref{fig:dep.coeff.Fe1} to \ref{fig:dep.coeff.Mg2} show the departure coefficients we obtained for Fe{\sc i}, Fe{\sc ii}, Fe{\sc iii}, and Mg{\sc i}, for the first 80 energy levels, and for Mg{\sc ii} for the first 20 energy levels (those included in the 17.02 Cloudy distribution). Throughout the upper atmosphere, the considered Fe{\sc i} energy levels are on average underpopulated, with the population increasing with pressure. We remark that the Fe{\sc i} underpopulation, particularly of the upper energy levels, might partially be an artifact of an incomplete Fe{\sc i} model atom, although for each element we considered all energy levels available in the 17.02 Cloudy distribution. In fact, \citet{mashonkina2011} showed that for late-type stars a more complete Fe{\sc i} model atom facilitates recombination from Fe{\sc ii}, decreasing the underpopulation of the upper levels of Fe{\sc i}.

The first $\approx$60 energy levels of Fe{\sc ii} are systematically overpopulated by about a factor of ten at pressures lower than $\approx$10$^{-5}$\,bar. Higher Fe{\sc ii} energy levels are instead slightly underpopulated. Figure~\ref{fig:dep.coeff.Fe2} shows also that at pressures higher than about 10$^{-5}$\,bar, Fe{\sc ii} can be treated in LTE. The Fe{\sc iii} departure coefficients behave similarly to those of Fe{\sc i}, with an average moderate underpopulation and with the level population increasing with pressure. At pressures higher than $\approx$10$^{-8}$\,bar, Fe{\sc iii} can be treated in LTE. Figures~\ref{fig:dep.coeff.Mg1} and \ref{fig:dep.coeff.Mg2} show that the considered Mg{\sc i} and Mg{\sc ii} energy levels are significantly underpopulated.

Therefore, the overpopulation of the lower levels of Fe{\sc ii} and the underpopulation of the lower levels of Mg{\sc i} and Mg{\sc ii} is the likely cause of the significant difference between the LTE and NLTE TP profiles, and thus also of the difference between the Cloudy NLTE TP profile and the HELIOS and PHOENIX TP profiles that have been computed assuming LTE. The importance of Fe{\sc ii} overpopulation in the atmospheric heating increases when considering that Fe{\sc ii} is the dominant Fe species at pressures between 10$^{-2}$ and 10$^{-10}$\,bar (Figure~\ref{fig:chemistry}), thus in the pressure range in which the temperature increases most significantly and reaches its maximum. Furthermore, the lines rising from the lower, overpopulated energy levels lie in the (near) ultraviolet, that is the wavelength band in which the stellar radiation is most intense.
\subsection{Importance of NLTE effects}\label{sec:importanceNLTE}
We further explored the importance of accounting for NLTE effects by computing the planetary transmission spectra in both LTE and NLTE. To compute the transmission spectrum in LTE, we first constructed a composite LTE TP profile by joining the Cloudy LTE TP profile in the upper atmosphere with the HELIOS TP profile in the lower atmosphere, as described in Section~\ref{sec:modelling}. In the case of the LTE Cloudy TP profile, we found that the factor scaling the stellar flux and leading to best match the HELIOS TP profile around the 10$^{-4}$\,bar pressure level is 0.67, which is very close to the HELIOS (i.e. LTE) $f$ value (Section~\ref{sec:lower}). To improve the accuracy in the calculation of the transmission spectra, particularly in setting the continuum level (see below), we resampled the composite LTE and NLTE TP profiles in the 1--10$^{-11}$\,bar pressure range and divided it into 50 layers equally spaced in $\log{p}$.

We computed the transmission spectra employing Cloudy and the algorithm described in \citet{young2020a} and \citet{fossati2020}. To obtain the radius profile of the atmosphere (Eq.~(\ref{eq.radius_H_P})), we determined the reference pressure--radius reference point ($p_0$ and $R_0$) by fitting iteratively for the reference pressure $p_0$ constrained by the observed transit radius considering the KELT filter bandpass, which barycenter lies at about 6000\,\AA\ ($R_0$\,=\,1.936\,$R_{\rm Jup}$\,=\,$R_{\rm p}$). It is necessary to perform again the $p_0$--$R_0$ calibration, because of the different geometry involved in computing transmission spectra, compared to what we considered for computing the TP profile described in Section~\ref{sec:upper} (i.e. from ``emission geometry'' to ``transmission geometry''). Following the procedure described in \citet{fossati2020}, we determined $p_0$ in an iterative procedure by integrating the transmission model spectrum over the KELT bandpass filter, adjusting $p_0$ until the band-integrated model planetary radius matches the observed transit radius of $R_{\rm p}$\,=\,1.936\,$R_{\rm Jup}$. In particular, at each iteration we determined $p_0$ looking for the pressure at which the optical depth integrated over the KELT bandpass filter reaches 0.56 \citep{lecavelier2008}. In this way, following two iterations, we obtained that $p_0$ lies at a pressure of 0.059 and 0.028\,bar for the NLTE and LTE cases, respectively. We obtained two different values of $p_0$ in LTE and NLTE, because the continuum optical depth depends on the entire photosphere that is being probed, which have different physical characteristics in LTE and NLTE. Furthermore, the two $p_0$ values have been obtained by computing the radiative transfer in different ways (i.e. assuming LTE or accounting for NLTE effects).

We remark that we assumed that the reference radius $R_0$ corresponds to the transit radius when computing both the TP profile at the substellar point and the transmission spectrum. However, this assumption has a negligible impact on the results. In contrast, when computing the transmission spectrum, we make a further more coarse assumption, namely that the TP profile computed at the substellar point is valid across the entire planet \citep[i.e. both day and night side; see ][for a discussion about this assumption for KELT-9b]{fossati2020}. As described in Section~\ref{sec:comparison}, it is possible that this assumption has an non-negligible impact on the line shapes in the transmission spectra.
\begin{figure}[h!]
\includegraphics[width=\hsize,clip]{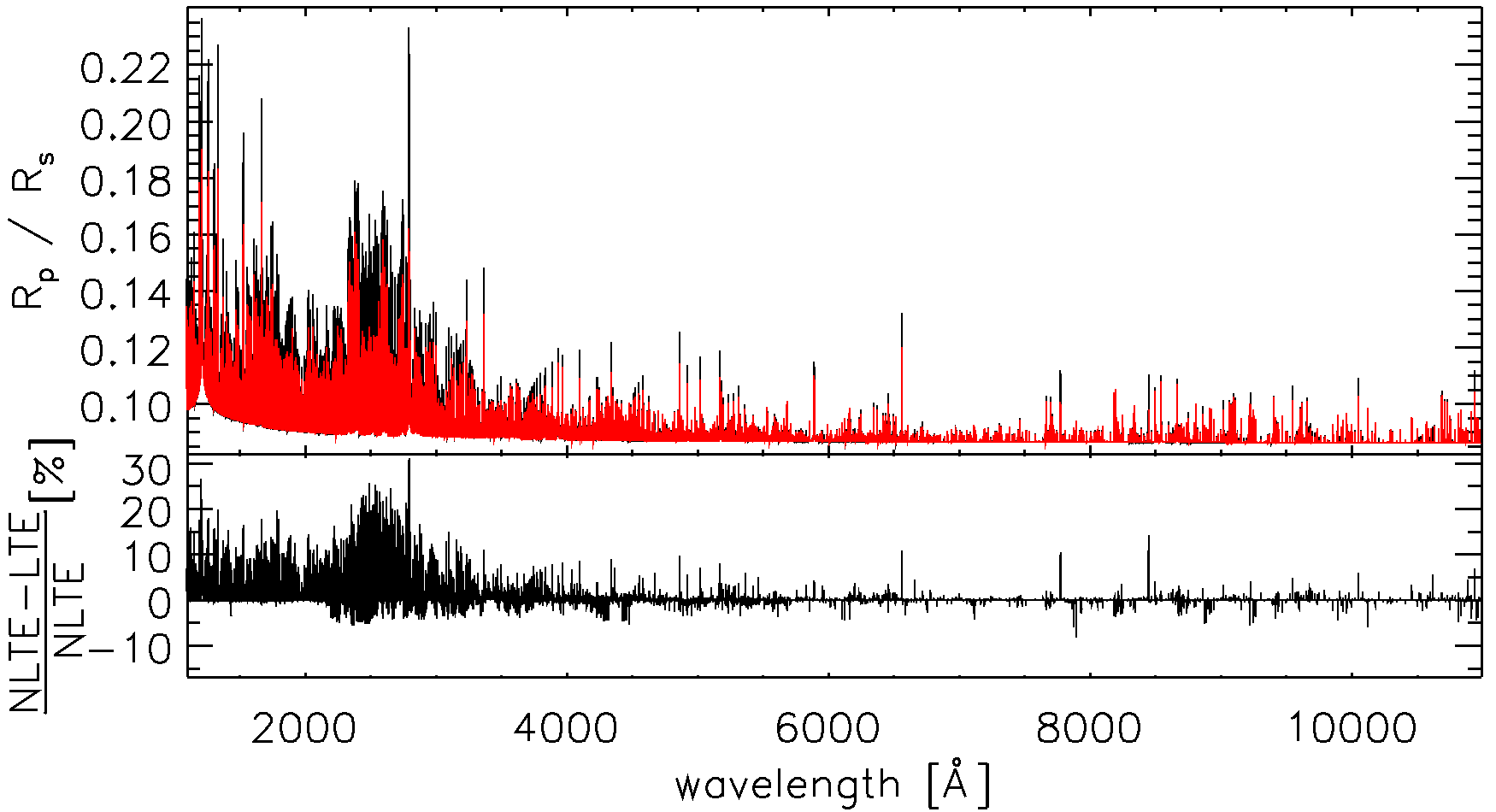}
\includegraphics[width=\hsize,clip]{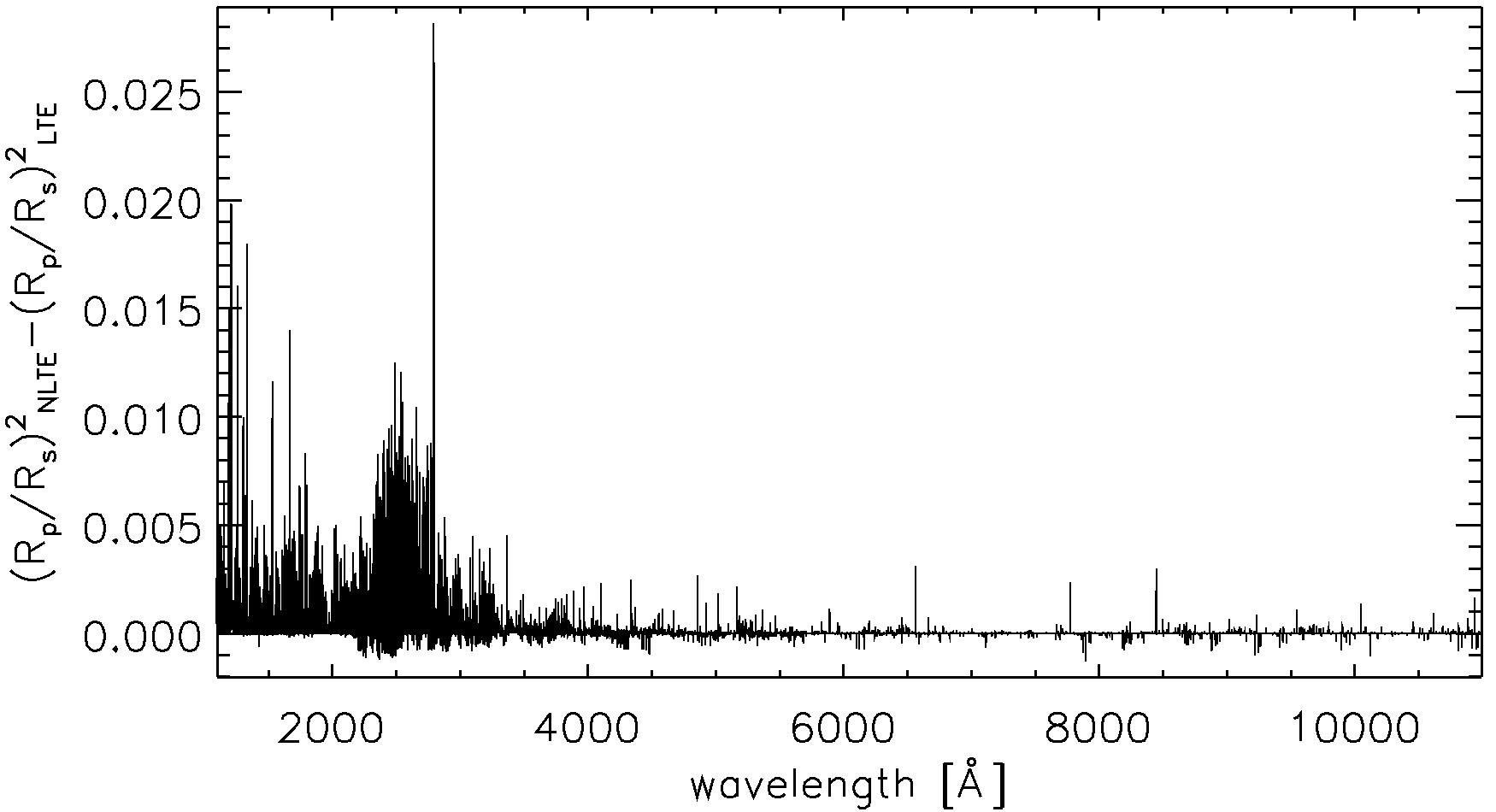}
\caption{Top: NLTE (black) and LTE (red) transmission spectra of KELT-9b ranging between the far ultraviolet and the near infrared. The transmission spectra have been computed considering a spectral resolution of 100\,000. The NLTE transmission spectrum has been computed on the basis of the NLTE TP profile, while the LTE transmission spectrum has been computed on the basis of the LTE TP profile. The bottom panel shows the deviation from LTE in \%. The strongest deviation from LTE is found at ultraviolet wavelengths. Bottom: Transit depth difference between the NLTE and LTE transmission spectra.}
\label{fig:transmission_all}
\end{figure}

Figure~\ref{fig:transmission_all} presents the LTE and NLTE transmission spectra in the 1100--11000\,\AA\ range, the NLTE correction, and the transit depth difference as a function of wavelength. Similar plots, but zooming into shorter wavelength ranges for better visibility can be found in Appendix~\ref{sec:appendix_transmissionSpectra}. Throughout the considered wavelength range, the LTE assumption leads mostly to underestimated absorption-line strengths, while the opposite occurs for a much smaller number of features. NLTE effects in the transmission spectrum are particularly strong in the ultraviolet wavelength range, where the deviation from LTE is on average well above 5--10\%, with a peak of about 30\% for the Mg{\sc ii}\,h\&k resonance lines. In the optical, instead, the deviation from LTE is smaller for most lines, although the deviation is as much as 10--15\% for a few specific features (e.g. H$\alpha$, O{\sc i} infrared triplet).

\begin{figure}[h!]
\includegraphics[width=\hsize,clip]{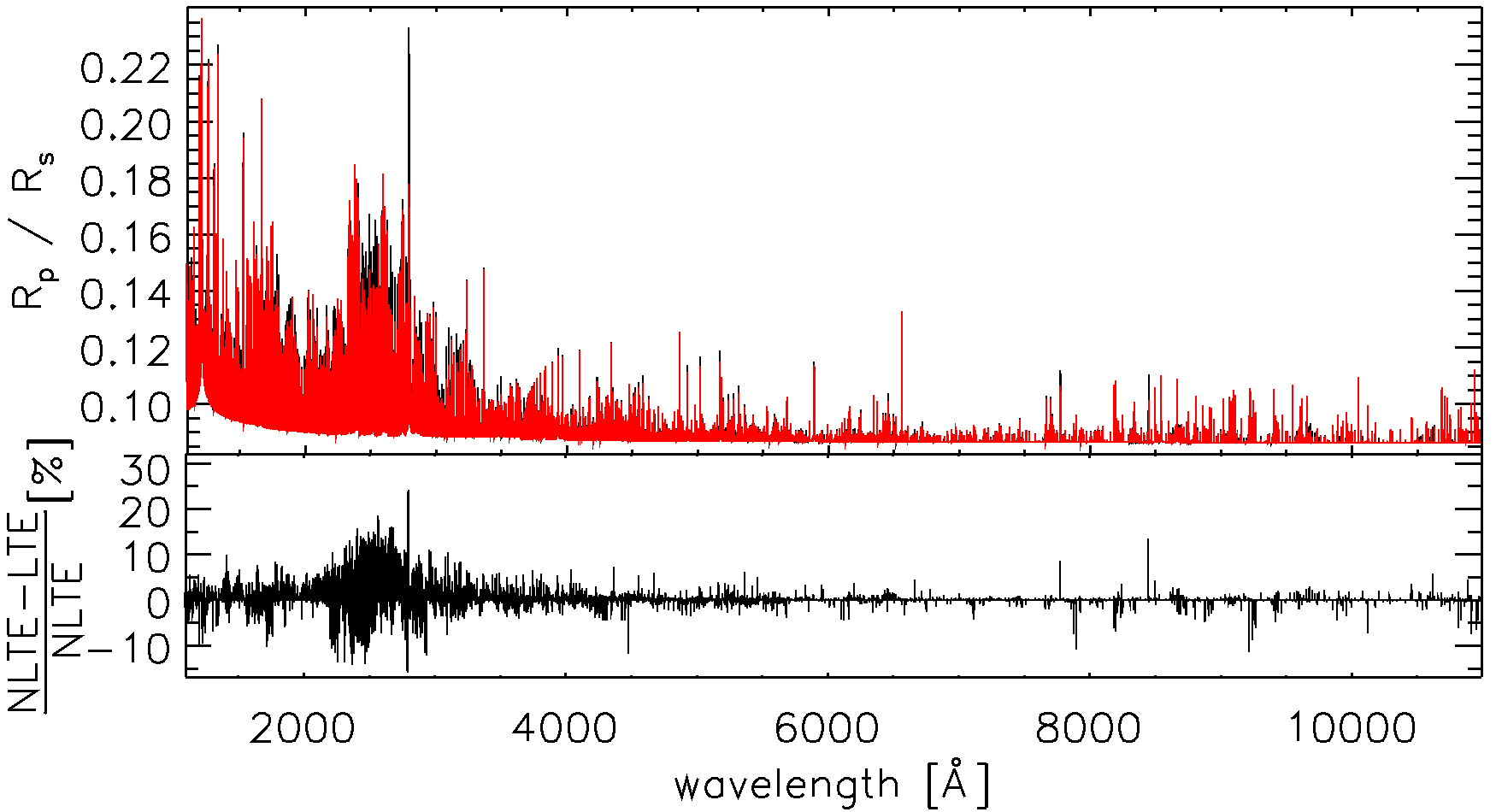}
\caption{Same as Figure~\ref{fig:transmission_all}, but comparing the transmission spectra computed in LTE (red) and NLTE (black) on the basis of the NLTE TP profile.}
\label{fig:transmission_mixed}
\end{figure}

A very significant contribution to the strong deviations from LTE, particularly at ultraviolet wavelengths, is given by the large temperature difference between the underlying LTE and NLTE TP profiles. The majority of the spectral lines at short wavelengths come from ionised species, which are more abundant in the NLTE model as a consequence of the generally higher atmospheric temperature of the NLTE TP profile compared to the LTE TP profile, particularly in the line forming region. Furthermore, the difference in the LTE and NLTE TP profiles leads to different scale heights, which also affect the features in the transmission spectrum. Figure~\ref{fig:transmission_mixed} demonstrates this by showing the comparison between the transmission spectra computed in LTE and NLTE on the basis of the NLTE TP profile. From Figures~\ref{fig:transmission_all} and \ref{fig:transmission_mixed} one can notice that, for example, for the H$\alpha$ line the NLTE correction has moved from about 10\% to zero. This is because Cloudy is not set up to do LTE calculations involving the hydrogen $n$\,=\,2 level (i.e. Lyman and Balmer lines) and therefore the NLTE correction for these lines shown in Figure~\ref{fig:transmission_all} is exclusively due to the difference between the LTE and NLTE TP profiles. 

In the far-ultraviolet (Figure~\ref{fig:transmission_lteVSnlte_ranges1}, top), the Si{\sc iii} resonance line at 1206\,\AA\ is the one showing the strongest deviation from LTE ($\sim$27\%). Several other far-ultraviolet features (e.g. Ly$\alpha$, C{\sc ii} 1335\,\AA, Si{\sc ii} 1530\,\AA, C{\sc iv} 1545\,\AA) present deviations from LTE above 15\%. Accounting for NLTE effects, we predict that several features in the planetary transmission spectrum will reach the 2--5\% absorption level, thus presenting transit light curves 2 to 6 times deeper than the continuum level, which would be most likely detectable with HST. In particular, silicon is a species that, though roughly as abundant as iron (assuming solar composition), has not yet been detected in the transmission spectrum of KELT-9b. The strong features at $\approx$1530\,\AA\ belong to Si{\sc ii} resonance lines and would be ideal targets for detecting Si in the planetary atmosphere.

The mid-ultraviolet spectral region (Figure~\ref{fig:transmission_lteVSnlte_ranges1}, bottom) is characterised by a forest of rather strong metal lines for which the NLTE correction is on average of the order of 10\%. However, there is a strong feature at $\approx$1670\,\AA\ that is a blend composed by two strong Fe{\sc ii} lines rising from low energy levels (i.e. 0.232 and 0.352\,eV) and a strong Al{\sc ii} resonance line. This strong blend presents a NLTE correction larger than 15\%.

In the near ultraviolet (Figure~\ref{fig:transmission_lteVSnlte_ranges2}), the strongest features are the Fe{\sc ii} absorption bands in the 2400--2800\,\AA\ wavelength range and the Mg{\sc ii}\,h\&k resonance lines at $\approx$2800\,\AA. These are strong features in both the LTE and NLTE transmission spectra, but they appear to be significantly enhanced when accounting for NLTE effects, with deviations from LTE consistently above 15\%. Interestingly, the NLTE transmission spectrum indicates that both Fe{\sc i} and Fe{\sc ii} near ultraviolet features are going to be particularly strong. This gives the opportunity to employ near ultraviolet transmission spectroscopy to detect both ions and thus observationally constrain the atmospheric Fe ionisation fraction, which would give additional precious constraints to the shape of the TP profile. The NLTE correction for the Mg{\sc ii}\,h\&k resonance lines reaches 30\%. This large correction is due to the difference in the temperature structure between the LTE and NLTE TP profiles. As a result of the fact that Mg ionises in the lower atmosphere, between the 10$^{-2}$ and 10$^{-3}$\,bar level, the Mg{\sc i} resonance line at 2853\,\AA\ is not a prominent feature in the planetary transmission spectrum, similarly to what observed for the ultra-hot Jupiter WASP-121b \citep{sing2019}, and reaches a transit depth of about 1.5\%.

The transmission spectrum in the blue part of the optical wavelength range (Figure~\ref{fig:transmission_lteVSnlte_ranges2}, bottom; Figure~\ref{fig:transmission_lteVSnlte_ranges3}, top) is characterised by several metal lines that however reach only about the 1--1.5\% absorption depth. There are however two features that peak significantly over the others reaching 2\% transit depths. These are strong Ti{\sc ii} lines rising from excited states. The Ca{\sc ii}\,H\&K resonance lines are only slightly stronger than the high order hydrogen Balmer lines and produce absorption depths in transmission of the order of 1.2\%. This part of the transmission spectrum also shows the presence of a large number of lines belonging mostly to Fe-peak elements, which have been detected both directly \citep[e.g.][]{cauley2019} and through cross-correlation \citep[e.g.][]{hoeijmakers2018,hoeijmakers2019}. 

In the red part of the optical wavelength range (Figure~\ref{fig:transmission_lteVSnlte_ranges3}, bottom), the H$\alpha$ line is the strongest feature in the transmission spectrum, reaching a depth of almost 1.5\%. \citet{fossati2020} found that the LTE assumption would lead to overestimate the strength of the hydrogen Balmer lines, while Figure~\ref{fig:transmission_all} shows the opposite. This is because the LTE transmission spectrum presented here is based on the LTE TP profile, which is significantly cooler than the NLTE TP profile, thus leading to a smaller concentration of excited hydrogen atoms and thus to smaller Balmer features. Despite the fact that Na is ionised in the lower atmosphere, there is enough Na{\sc i} in the line forming region to produce quite prominent Na{\sc i}D features that reach an absorption depth in transmission of about 1\%. Interestingly, the feature rising to above the 1\% level absorption depth at about 7780\,\AA\ is the O{\sc i} near-infrared triplet that could be detectable in the already obtained high-resolution transmission observations.

Figure~\ref{fig:transmission_all}, and in particular the more detailed plots in Appendix~\ref{sec:appendix_transmissionSpectra}, show also that the optical and near-infrared wavelength ranges host a number of lines with negative NLTE correction (i.e. the line is stronger in LTE than in NLTE). The most prominent of these lines with negative NLTE corrections $\gtrsim$2.5\% lie between 2200 and 2600\,\AA\ and at $\approx$4300, 4450, and 7900\,\AA. The lines with negative NLTE corrections in the near-ultraviolet wavelength range belong primarily to features of Fe-peak elements (mostly neutral) rising from underpopulated energy levels. The lines at $\approx$4300\,\AA\ and at wavelengths shorter than 4460\,\AA\ are all strong Ca{\sc i} features rising from levels with an energy of about 1.88\,eV. The features at $\approx$4480 and 7900\,\AA\ correspond to the Mg{\sc ii} triplets at 4481 and 7890\,\AA.

We leave the detailed analysis of the NLTE transmission spectrum and its comparison to the available observations, particularly of metal lines, to a future work. However, we check the quality of the NLTE TP profile and transmission spectrum by comparing the synthetic profiles of the H$\alpha$ and H$\beta$ lines with the observations. This comparison is the focus of the next section.

Figure~\ref{fig:transmission_all} and the above discussion demonstrate that NLTE effects play a fundamental role in shaping not only the planetary atmospheric TP profile, but also the transmission spectrum. The large difference between the predicted LTE and NLTE transmission spectra for several spectral features gives the further opportunity to observationally test whether the atmospheric temperature is indeed as high as predicted by our NLTE calculations and thus confirm or disprove the impact of NLTE effects in driving the atmospheric properties of ultra-hot Jupiters.
\subsection{Comparison with observations}\label{sec:comparison}
We compare here the synthetic LTE and NLTE transmission spectrum of the H$\alpha$ and H$\beta$ lines with the available observations. In particular, we consider the observations presented by \citet[][H$\alpha$]{yan2018}, \citet[][H$\alpha$ and H$\beta$]{cauley2019}, \citet[][H$\alpha$]{turner2020}, and \citet[][H$\alpha$ and H$\beta$]{wyttenbach2020}. We refer to Section~2 of \citet{fossati2020} for a thorough discussion of the similarities and differences, also in terms of data analysis, present among these observations.
\begin{figure}[h!]
\includegraphics[width=\hsize,clip]{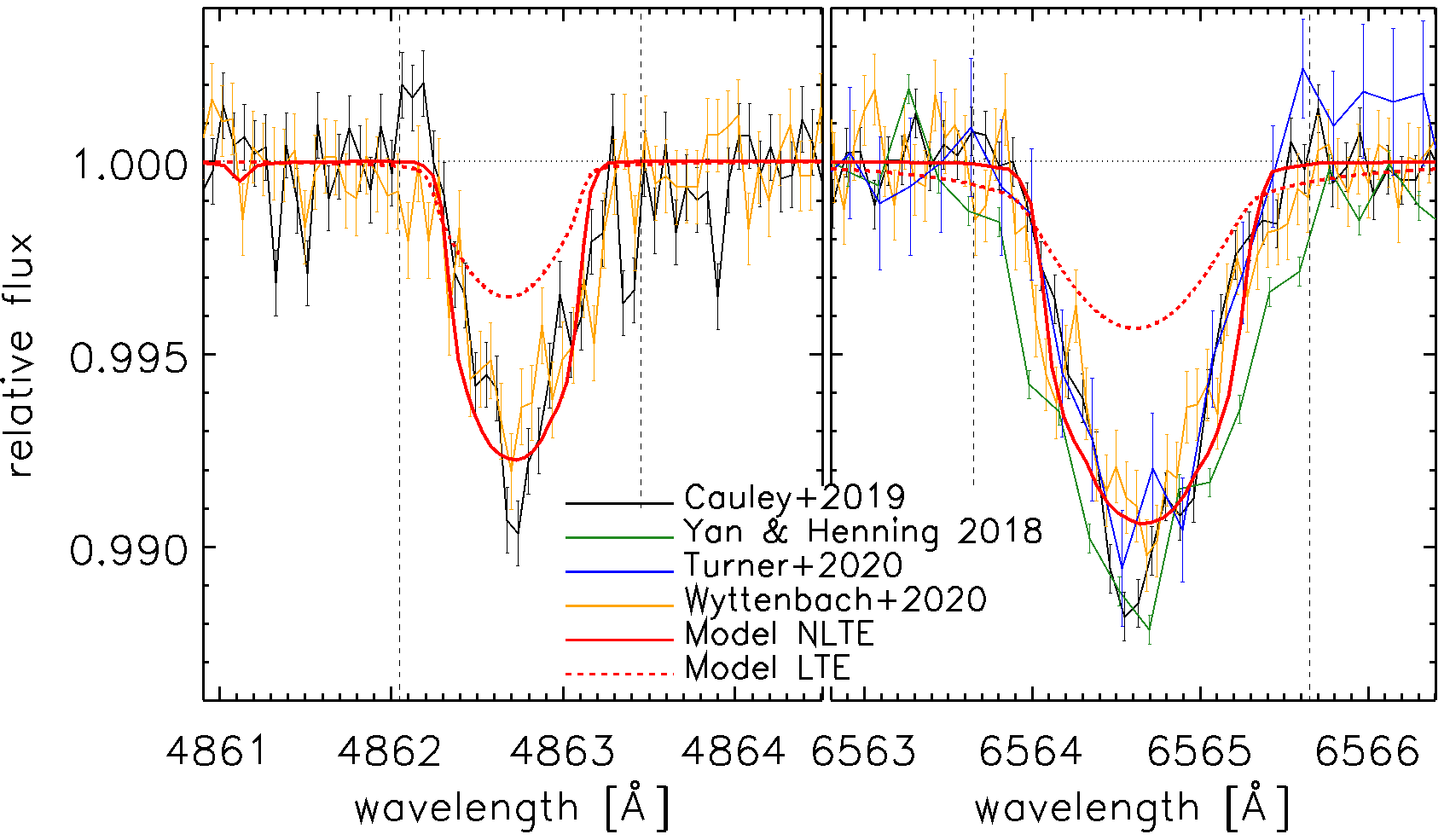}
\caption{Transmission spectra of the H$\alpha$ (right) and H$\beta$ (left) lines presented by \citet[][green]{yan2018}, \citet[][black]{cauley2019}, \citet[][blue]{turner2020}, and \citet[][orange]{wyttenbach2020}. Wavelengths are in vacuum to match those of the synthetic transmission spectrum. The data have been rebinned by a factor of six for visualisation purposes and have been aligned to the same wavelength employing Gaussian fits \citep[see][]{fossati2020}. The thick red solid and dashed lines show the synthetic NLTE and LTE transmission spectra, respectively. The black, vertical dashed lines enclose the wavelength ranges considered to compute the $\chi^2$ and $\chi^2_{\rm red}$ values. The horizontal dotted line at one sets the continuum level to guide the eye.}
\label{fig:comparisonHaHb}
\end{figure}

Figure~\ref{fig:comparisonHaHb} shows the comparison between the observed and synthetic NLTE and LTE H$\alpha$ and H$\beta$ line profiles. The LTE transmission spectrum has been computed employing the LTE TP profile (Figure~\ref{fig:lteVSnlte}) and the procedure described in Section 6.3 of \citet{fossati2020}. This is because, by construction, Cloudy is not set to compute the $n$\,=\,1 and $n$\,=\,2 hydrogen level populations in LTE. For context, Figure~\ref{fig:hydrogen_b} shows the H{\sc i} departure coefficients for the first ten energy levels, showing that the lower three energy levels (responsible for the Lyman, Balmer, and Paschen series) are heavily overpopulated in the upper atmosphere.

The NLTE synthetic spectra of both lines are in general a good match to the data, while the LTE synthetic spectra are significantly weaker than the observations, mostly due to the cooler TP profile compared to the NLTE case. In the following, we consider only the NLTE synthetic profiles. 

The main difference between the synthetic spectra and the observations is the line shape, where the observed line profiles have a more triangular shape in comparison to the rounder shape of the synthetic lines. This difference might be due to the fact that we consider only the day-side temperature profile to compute the transmission spectra \citep[see also Section 6.7 of][]{fossati2020}. Indeed, the line profiles forming in the planetary night side would be narrower, because of the lower temperature and thus they would contribute mostly to the shape of the line core. We quantify the fit of the NLTE synthetic spectrum by computing the $\chi^2$ and reduced $\chi^2$ ($\chi^2_{\rm red}$) employing the non-rebinned observed spectra. The values are listed in Table~\ref{tab:chi2}.
\begin{table}[ht!]
\caption{$\chi^2$ and reduced $\chi^2$ values obtained from the comparison between the synthetic NLTE transmission spectrum and observations.}
\label{tab:chi2}
\begin{center}
\begin{tabular}{l|cc|cc}
\hline
\hline
       & \multicolumn{2}{c|}{H$\alpha$} & \multicolumn{2}{c}{H$\beta$} \\
Source & $\chi^2$ & $\chi^2_{\rm red}$ & $\chi^2$ & $\chi^2_{\rm red}$\\
\hline
\citet{yan2018}        & 497.46 & 7.54 &        &      \\
\citet{cauley2019}     & 159.78 & 1.11 & 159.16 & 1.17 \\
\citet{turner2020}     &  43.34 & 0.66 &        &      \\
\citet{wyttenbach2020} & 184.63 & 0.93 & 138.62 & 1.00 \\
\hline
\end{tabular}
\end{center}
\end{table}

The worst match is that with the H$\alpha$ profile of \citet{yan2018}, because the NLTE synthetic profile is significantly weaker and narrower than the observation. Indeed, the H$\alpha$ profile of \citet{yan2018} is by far the strongest of the four \citep[i.e. its equivalent width is about 10$\sigma$ larger than those of][]{cauley2019,wyttenbach2020}. The rather low $\chi^2$ and $\chi^2_{\rm red}$ values obtained from the comparison with the H$\alpha$ profile of \citet{turner2020} are driven by the large observational uncertainties, though visual inspection suggests that the synthetic spectrum is a good match to the data. 

The profiles of \citet{cauley2019} and \citet{wyttenbach2020} are the most comparable in terms of uncertainties and line strength \citep[][Table~1]{fossati2020}, though the profiles of \citet{cauley2019} are slightly deeper and narrower than those of \citet{wyttenbach2020}. The $\chi^2$ and $\chi^2_{\rm red}$ values listed in Table~\ref{tab:chi2} indicate that the NLTE transmission spectrum is a good match to these observations. Therefore, accounting for metals and NLTE effects in the computation of the TP profile and of the transmission spectrum produces a good fit to the observed hydrogen Balmer lines.

We remark that Cloudy is a static code, meaning that it does not account for hydrodynamic motions of the gas, but the model still leads to a good fit to the observed hydrogen Balmer lines. This is because the core of the H$\alpha$ and H$\beta$ lines probe about the nbar and 10\,nbar pressure levels\footnote{We arrived at this conclusion thanks to having solved the pressure-radius reference location degeneracy by matching the reference continuum pressure to the optical transit radius \citep{brogi2019,fisher2020}}, respectively, which are well below the sonic point (located between the 10$^{-11}$ and 10$^{-12}$\,bar), where hydrodynamic motions become significant. This indicates that it is not possible to use the fit of these lines to directly constrain hydrodynamic motions of the planetary atmosphere and thus mass-loss rates. However, the TP profile gives information regarding the energy available in the atmosphere to drive mass loss. This is going to be the subject of a future work.

The TP profile presented here and obtained through forward modelling appears to be a better fit to the data compared to what had been found through a grid approach by \citet{fossati2020}. The likely reason for this is that the algorithm used to compute the parametric TP profiles considered by \citet{fossati2020} did not provide enough flexibility to lead to a better match to the data compared to what had been obtained. In particular, none of the models in the grid of \citet{fossati2020} reached the maximum temperature at pressures as high as $\approx$10$^{-6}$\,bar, remaining then isothermal at lower pressures. Remarkably, this pressure level is at the center of the H$\alpha$ and H$\beta$ line formation region. In other words, the TP profiles employed by \citet{fossati2020} presenting a temperature at the top of the atmosphere of 8000-9000\,K were too cool around the main line formation region to lead to enough excited hydrogen atoms for matching the data. Indeed, the TP profile \citet{fossati2020} found to best match the data reached a temperature of 8000-8500\,K, the highest temperature in the forward model, around the 10$^{-6}$\,bar level (i.e. the center of the H$\alpha$ and H$\beta$ line formation region).

\begin{figure}[h!]
\includegraphics[width=\hsize,clip]{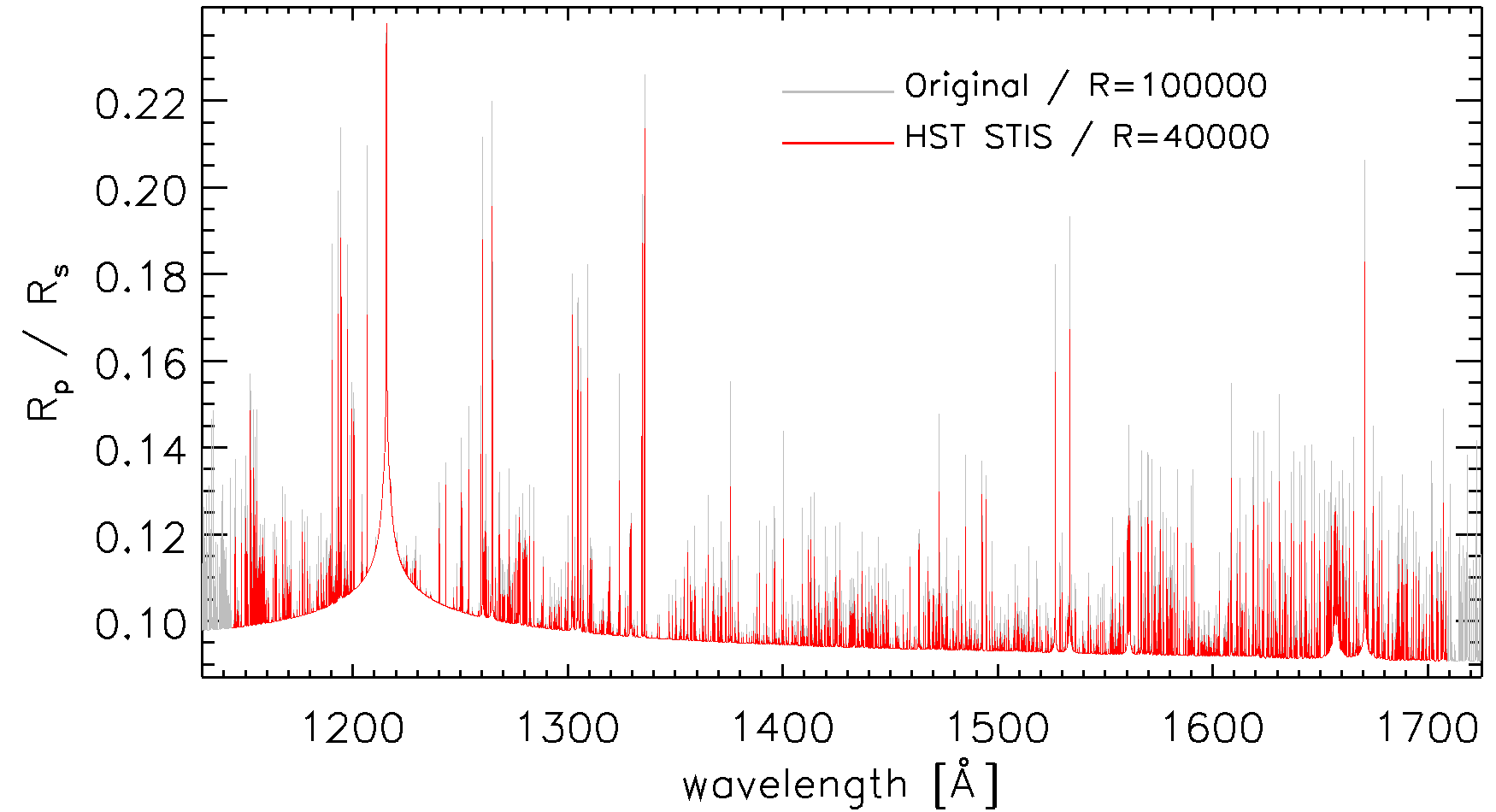}
\includegraphics[width=\hsize,clip]{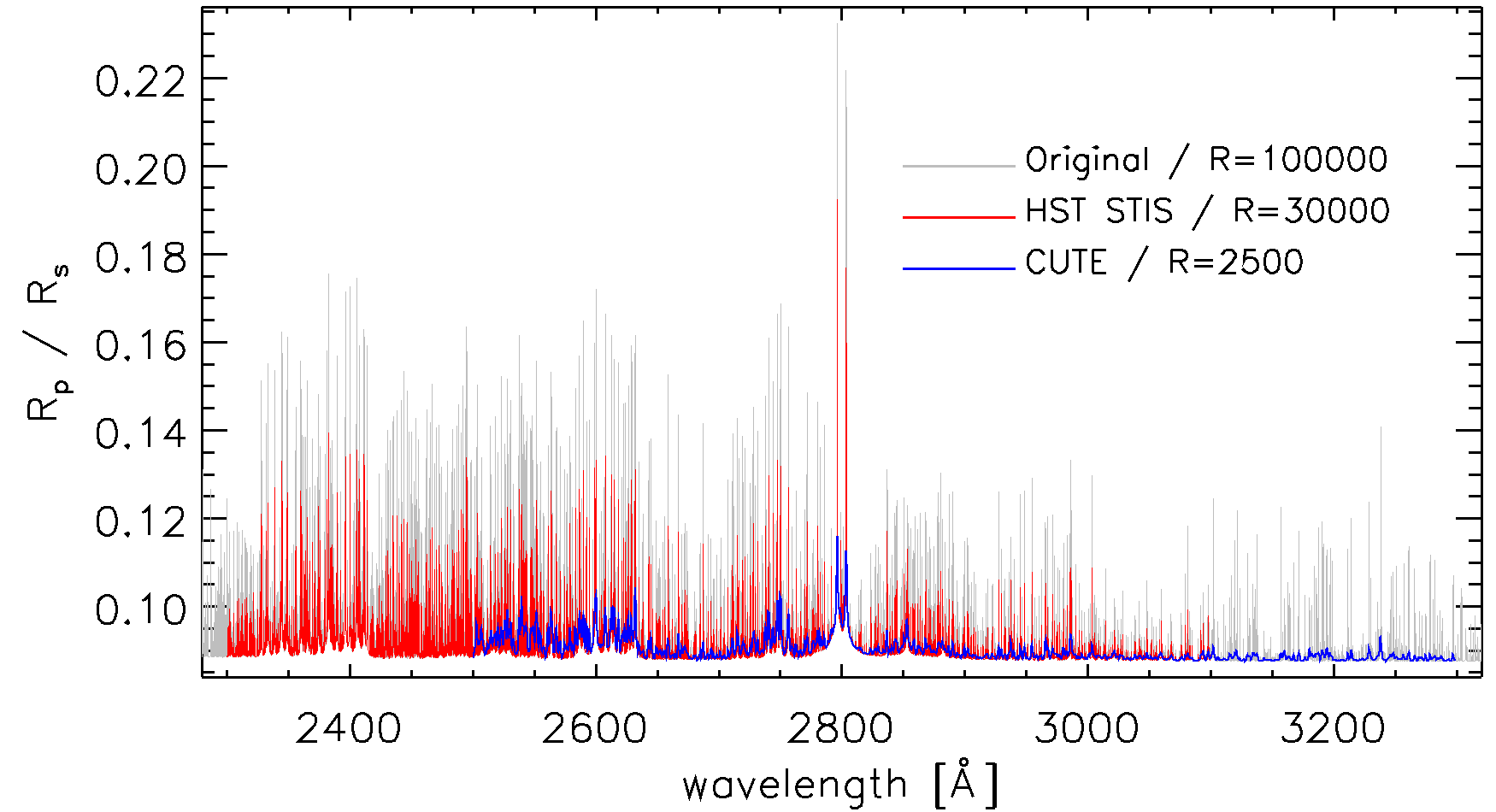}
\caption{Top: NLTE transmission spectrum in the far ultraviolet wavelength region (gray) convolved to the spectral resolution of the HST SITS E140M grating (red). Bottom: NLTE transmission spectrum in the far ultraviolet wavelength region (gray) convolved to the spectral resolution of the HST SITS E230M grating (red) and to that of the spectrograph on board the CUTE SmallSat mission (blue). The considered CUTE spectral resolution accounts for spacecraft jitter.}
\label{fig:uv}
\end{figure}

Finally, KELT-9b has been extensively observed at optical and near-infrared wavelengths, but ultraviolet transmission spectra are still not available. Furthermore, the ultraviolet is the wavelength range in which the stellar emission is highest. Therefore, to guide future ultraviolet observations, we convolved the NLTE transmission spectrum to the spectral resolution ($R$) of the HST SITS instrument in the far (E140M grating; 1140--1710\,\AA; $R$\,$\approx$\,40\,000) and near (E230M grating; 2280--3120\,\AA; $R$\,$\approx$\,30\,000) ultraviolet and of the spectrograph on-board of the CUTE SmallSat mission \citep[2500--3300\,\AA, $R$\,$\approx$\,2\,500; ][]{fleming2018}. The convolved transmission spectra are shown in Figure~\ref{fig:uv}.

We employed the HST/STIS and CUTE signal-to-noise (S/N) calculators to obtain the information necessary to generate simulated far ultraviolet and near ultraviolet transmission spectra collected with both instruments. The duration of KELT-9b's transit is of about 3.9 hours \citep{gaudi2017}, therefore it is possible to perform in-transit observations along two HST orbits, for a total of about 6000 seconds. Considering this exposure time, two transit observations, and the E140M and E230M gratings, we obtained average S/N values per pixel of about 85 and 310, respectively. For CUTE, instead, the S/N calculator, based on the CUTE data simulator \citep{sreejith2019}, indicates that the average S/N value per pixel obtained for KELT-9b following a 5 minutes exposure \citep[the baseline for CUTE observations;][]{fleming2018} is about 40. Assuming an observing efficiency of 65\% (i.e. accounting for Earth occultations, passes within the South Atlantic Anomaly, and readout time), we obtained that CUTE would be able to collect roughly 30 exposures of 5 minutes each, hence finally obtaining spectra with an average S/N per pixel, integrating over the whole transit, of about 220. The baseline for CUTE is to collect ten transits for each target, thus enabling to reach a final S/N per pixel of about 700. Figures~\ref{fig:simulated_observations_HST} and \ref{fig:simulated_observations_CUTE} show the transmission spectra obtained considering instrument spectral resolution, wavelength sampling, binning (ten pixels in the case of the HST/STIS observations and two pixels in the case of CUTE observations), and the S/N values given above. In the far ultraviolet spectral range, all major features corresponding to resonance lines of C{\sc ii}, Si{\sc ii}, and C{\sc iv} are detectable (i.e. consecutive datapoints at wavelengths corresponding to these lines lie above the noise level). In the near ultraviolet, both Mg{\sc ii}\,h\&k resonance lines and the strongest Fe features are detectable employing both HST/STIS and CUTE.
\section{Conclusion}\label{sec:conclusion}
We employed the Cloudy NLTE radiative transfer code to self-consistently compute the upper atmospheric TP profile of the prototype ultra-hot Jupiter KELT-9b, accounting for NLTE effects. We obtained a TP profile about 2000\,K hotter than predicted by other models assuming LTE. In particular, the profile displays a steep temperature rise in the 1--10$^{-7}$\,bar range, with the temperature increasing from $\approx$4000\,K to $\approx$8500\,K, remaining roughly constant at lower pressures. We run an additional Cloudy model for the upper atmosphere of KELT-9b, but assuming LTE, obtaining a TP profile comparable to those computed in LTE with other codes. Therefore, the stark difference in the upper atmospheric TP profile between our Cloudy NLTE model and those presented in the literature is due to NLTE effects. 

We examined in detail the output of the LTE and NLTE  Cloudy models, which indicate that Fe and Mg are responsible, respectively, for most of the upper atmospheric heating and cooling. We further explored this finding by computing new TP structure models with Cloudy in NLTE excluding Fe or Mg at a time. Following the removal of Fe from the atmospheric metal content, we obtained a TP profile significantly cooler than the original, and comparable to that computed in LTE. Instead, the removal of Mg from the atmospheric composition led to an upper atmospheric TP profile that is about 500\,K hotter than the original. The departure coefficients computed by Cloudy for both Fe and Mg indicate that in the upper atmosphere, most Fe{\sc i}, Fe{\sc iii}, Mg{\sc i}, and Mg{\sc ii} levels are underpopulated, while most Fe{\sc ii} levels are overpopulated. The Mg underpopulation leads to a lack of cooling compared to the LTE case. Instead, the Fe{\sc ii} overpopulation leads to an increased heating, particularly because the majority of the strong Fe{\sc ii} lines, rising from overpopulated energy levels, lie in the near ultraviolet wavelength range, which is the spectral band in which the host star's spectral energy distribution peaks. 

We further employed Cloudy to compute LTE and NLTE transmission spectra of KELT-9b covering from the far ultraviolet to the mid infrared. We found that the strongest deviations form LTE (up to 30\%) occur at ultraviolet wavelengths, though there are also features at longer wavelengths for which the NLTE correction is larger than 10\%. 

We compared the NLTE synthetic transmission spectrum with the observed H$\alpha$ and H$\beta$ lines obtaining an excellent match, and thus validating our results. This positive outcome hints at the fact that it could be possible to use transmission spectra computed accounting for NLTE effects to guide the search for spectral features in the already available and future transit observations of KELT-9b. To this end, we employed the NLTE synthetic transmission spectrum to compute synthetic ultraviolet observations obtained with HST/STIS, after two transits, and CUTE, after ten transits. The results indicate that the majority of the strong features, such as the C{\sc ii}, Si{\sc ii}, and C{\sc iv} resonance lines in the far ultraviolet and the Fe{\sc ii} bands and the Mg{\sc ii}\,h\&k resonance lines in the near ultraviolet would be detectable with both instruments. These are also the features presenting the strongest deviations from LTE, which would therefore provide a strong element for validating our results.

The striking difference between the LTE and NLTE TP profiles presented in this work suggests that, at least for KELT-9b, the LTE assumption might induce misinterpreting both observational and theoretical results. Furthermore, the strong temperature rise of about 2500\,K across the main line formation region invalidates the isothermal assumption that is often taken to simplify atmospheric modelling to enable retrieval analyses. Therefore, any result obtained on KELT-9b's atmosphere and based on an isothermal profile should be taken with caution, particularly if those have been obtained through comparisons with high-resolution observations.

Finally, it will be extremely important to test in the near future whether the importance of NLTE effects in controlling the atmospheric TP profile is a peculiarity of KELT-9b or it is a feature common also for less irradiated planets. Indeed, radiative transfer considerations suggest that NLTE effects might be even more relevant for cooler planets, because of the decreased importance of collisions. If this is the case, it will be critical to identify the parameter space within which NLTE significantly affects TP profiles. This work is needed also to adequately interpret the many observations of ultra-hot Jupiters that have been and are being obtained with several facilities, both from space and from the ground. Furthermore, most current models of giant planet upper atmospheres and escape do not include the metals that control the energy balance on ultra-hot Jupiters, such as KELT-9b. There is an urgent need to include the relevant species in these models to properly calculate mass loss rates and to interpret observations of metal lines lying at ultraviolet wavelengths that probe the uppermost layers of the atmosphere.
\begin{acknowledgements}
M.E.Y. acknowledges funding from the \"OAW-Innovationsfonds IF\_2017\_03. D.S. acknowledges the support of the DFG priority program SPP-1992 ``Exploring the Diversity of ExtrasolarPlanets'' (DFG PR 36 24 602/41). M.E.Y. acknowledges funding from the European Research Council (ERC) under the European Union's Horizon 2020 research and innovation program under grant agreement No 805445. A.G.S. and L.F. acknowledge financial support from the Austrian Forschungsf\"orderungsgesellschaft FFG project CONTROL P865968.
\end{acknowledgements}
\begin{appendix}
\section{Cloudy TP profiles computed considering different reference radii}\label{sec:appendix_TP_Rp}
%
\begin{figure}[h!]
\includegraphics[width=\hsize,clip]{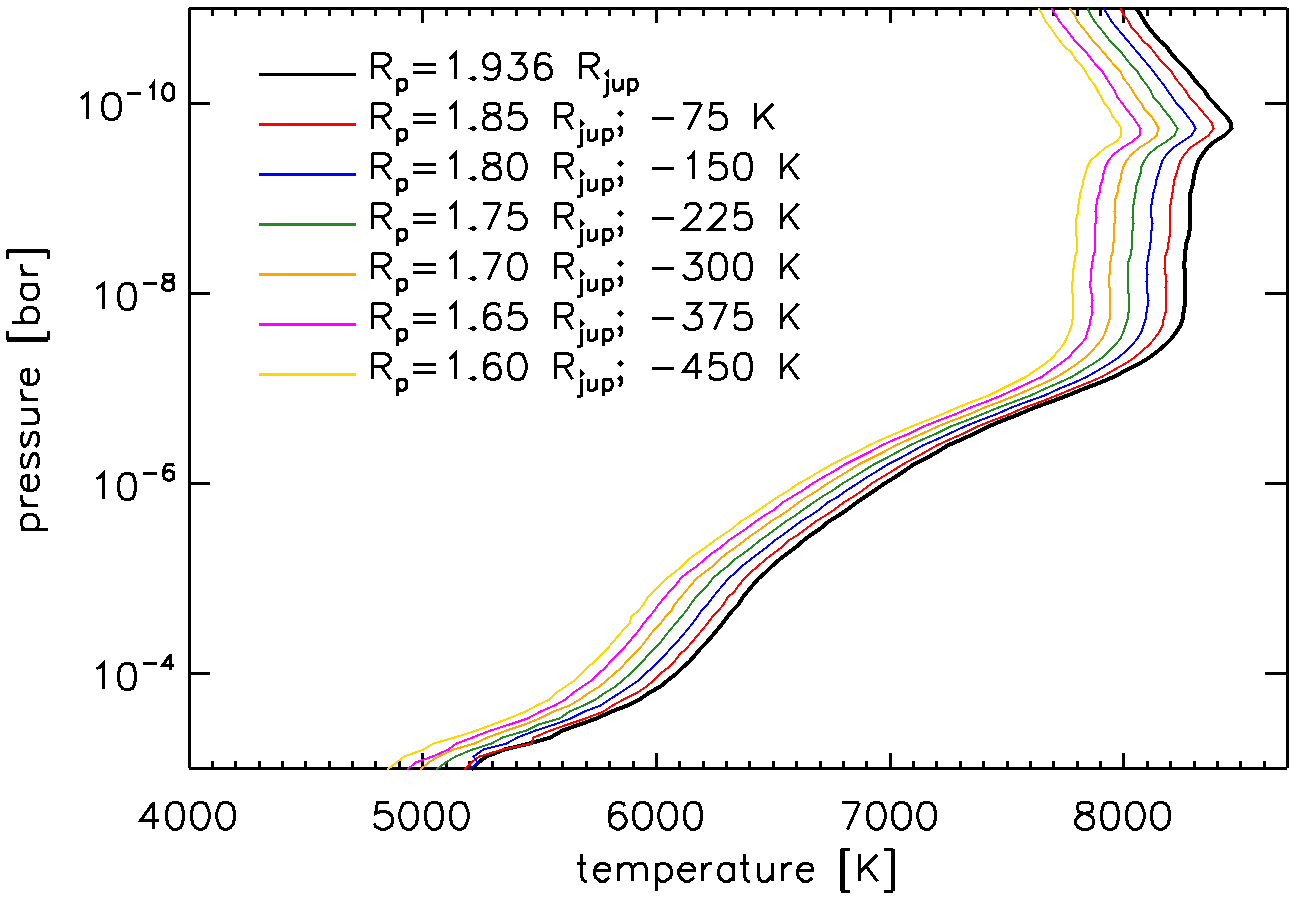}
\caption{Cloudy TP profiles computed considering different reference radii. The adopted TP profile is shown in black. For visualisation purposes, the TP profiles are rigidly shifted horizontally by the value indicated in the legend.}
\label{fig:TP_Rp}
\end{figure}
%
\section{Cloudy TP profiles computed considering different numbers of atmospheric layers}\label{sec:appendix_TP_profiles_layers}
%
\begin{figure}[h!]
\includegraphics[width=\hsize,clip]{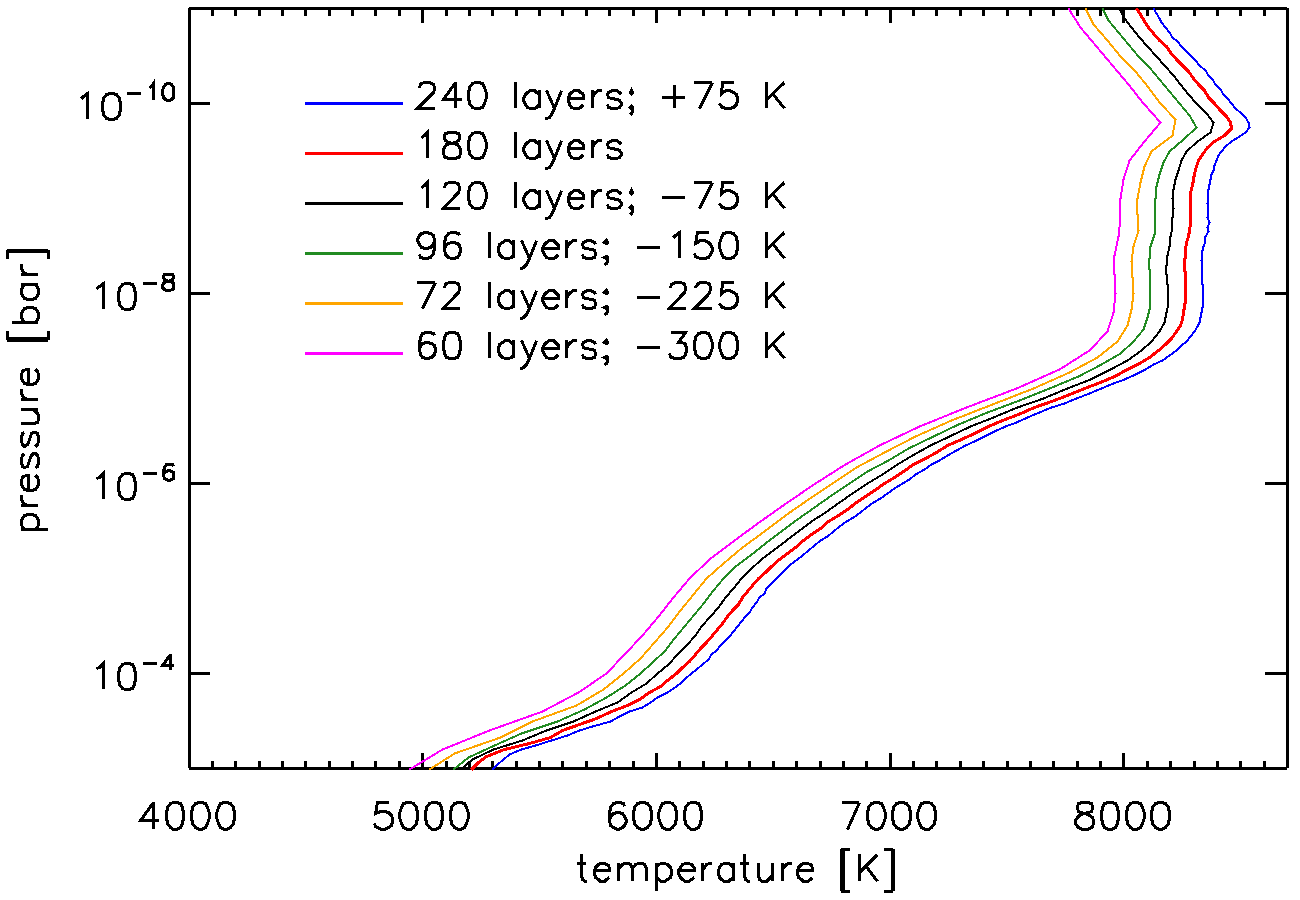}
\caption{Cloudy TP profiles computed considering different numbers of atmospheric layers. The finally adopted TP profile is shown in red. For visualisation purposes, the TP profiles are rigidly shifted horizontally by the value indicated in the legend.}
\label{fig:TPlayers}
\end{figure}
%
\section{Iron and magnesium departure coefficients}\label{sec:appendix_departure_coefficients}
%
\begin{figure*}[h!]
\includegraphics[width=\hsize,clip]{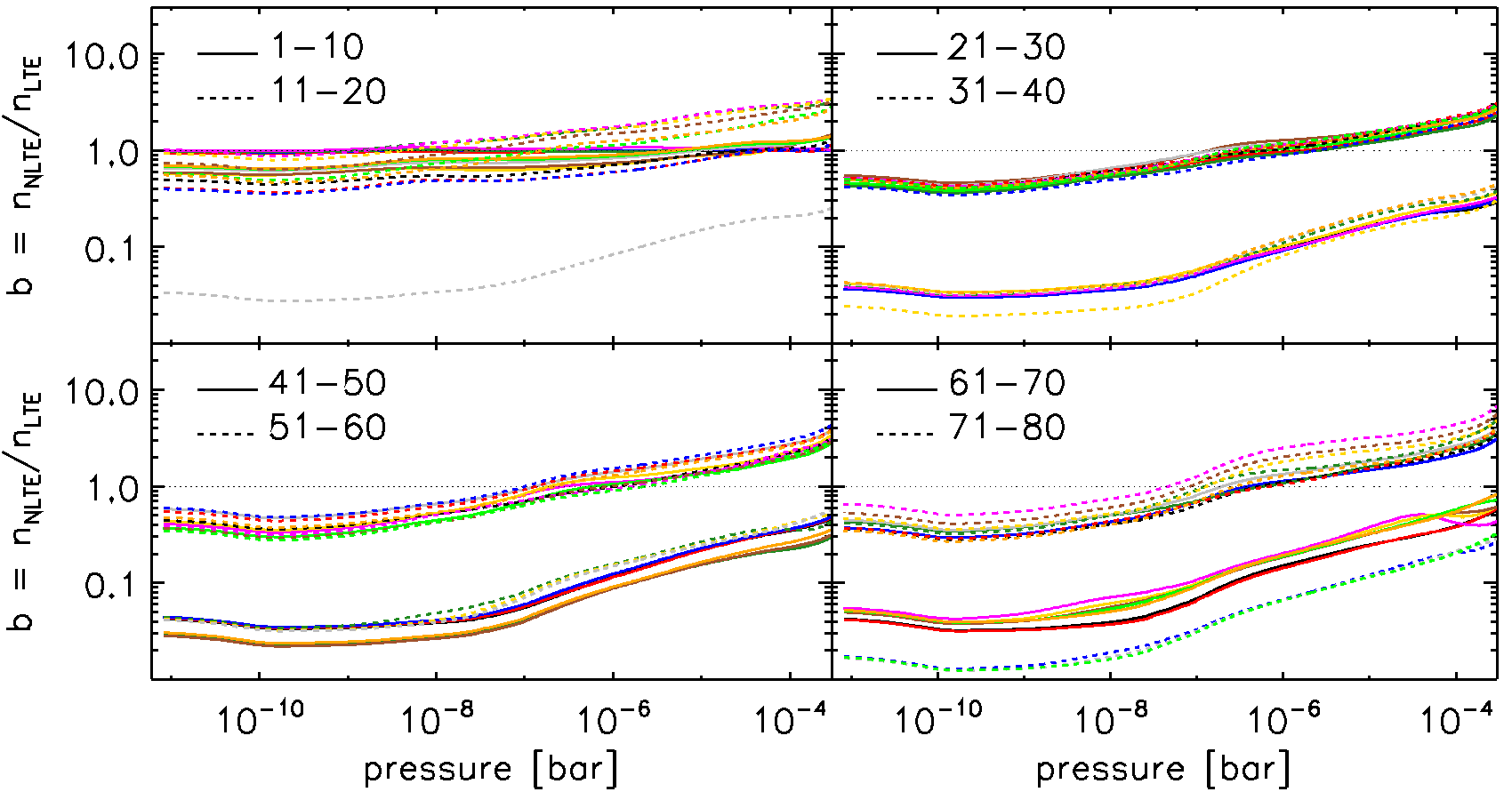}
\caption{Departure coefficients as a function of atmospheric pressure for the first 80 energy levels of Fe{\sc i}. The energy levels are numbered from 1 to 80 and are separated in groups of ten among the four panels and two line styles (solid and dashed) for each panel. Within each group of ten energy levels, the order of the line colors corresponding to increasing energy is black, red, blue, dark green, magenta, yellow, brown, gray, bright green, and orange. Therefore, the black solid line shows $b$ for the ground state (i.e. $n$\,=\,1 level), the red solid line shows $b$ for $n$\,=\,2 level, the blue solid line shows $b$ for the $n$\,=\,3 level, ..., the black dashed line shows $b$ for the $n$\,=\,11 level, etc. The dotted line at one indicating $n_{\rm NLTE}$\,=\,$n_{\rm LTE}$ is for reference. The departure coefficients that appear to be diverging from one with increasing pressure do reach one in the lower atmosphere, latest around the 0.2\,bar level, at which point all departure coefficients are equal to one.}
\label{fig:dep.coeff.Fe1}
\end{figure*}
\begin{figure*}[h!]
\includegraphics[width=\hsize,clip]{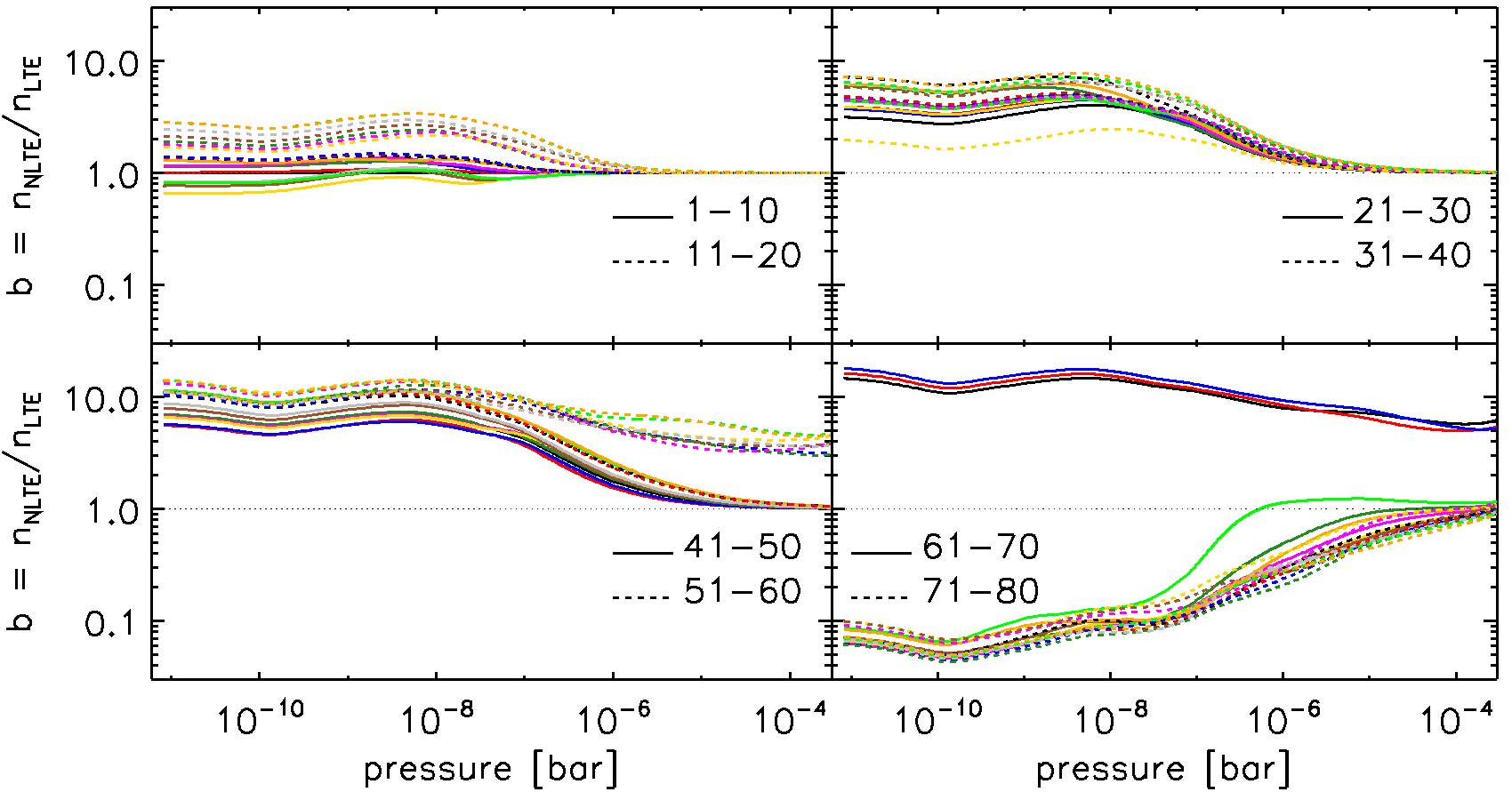}
\caption{Same as Figure~\ref{fig:dep.coeff.Fe1}, but for Fe{\sc ii}.}
\label{fig:dep.coeff.Fe2}
\end{figure*}
\begin{figure*}[h!]
\includegraphics[width=\hsize,clip]{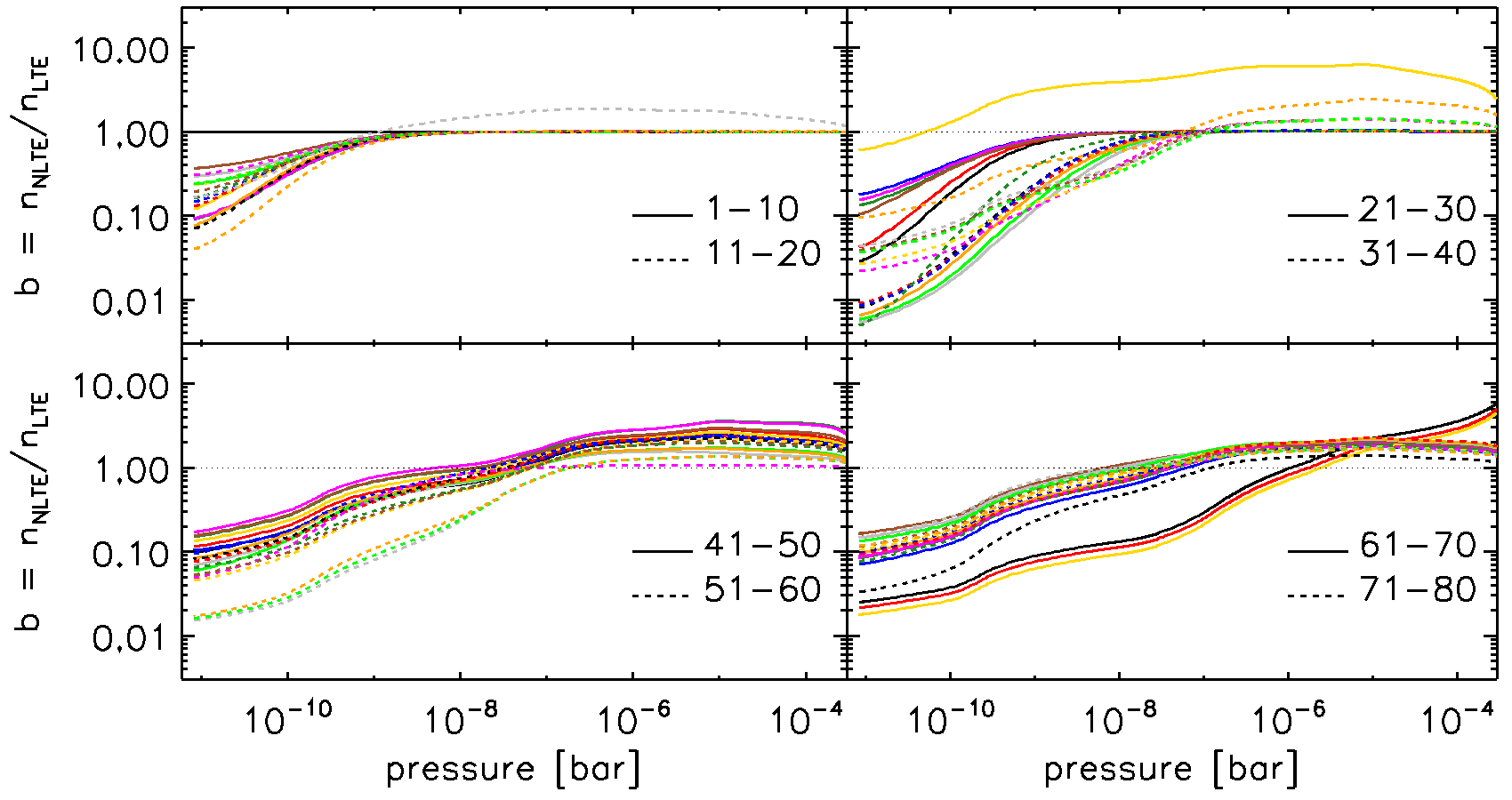}
\caption{Same as Figure~\ref{fig:dep.coeff.Fe1}, but for Fe{\sc iii}.}
\label{fig:dep.coeff.Fe3}
\end{figure*}
\begin{figure*}[h!]
\includegraphics[width=\hsize,clip]{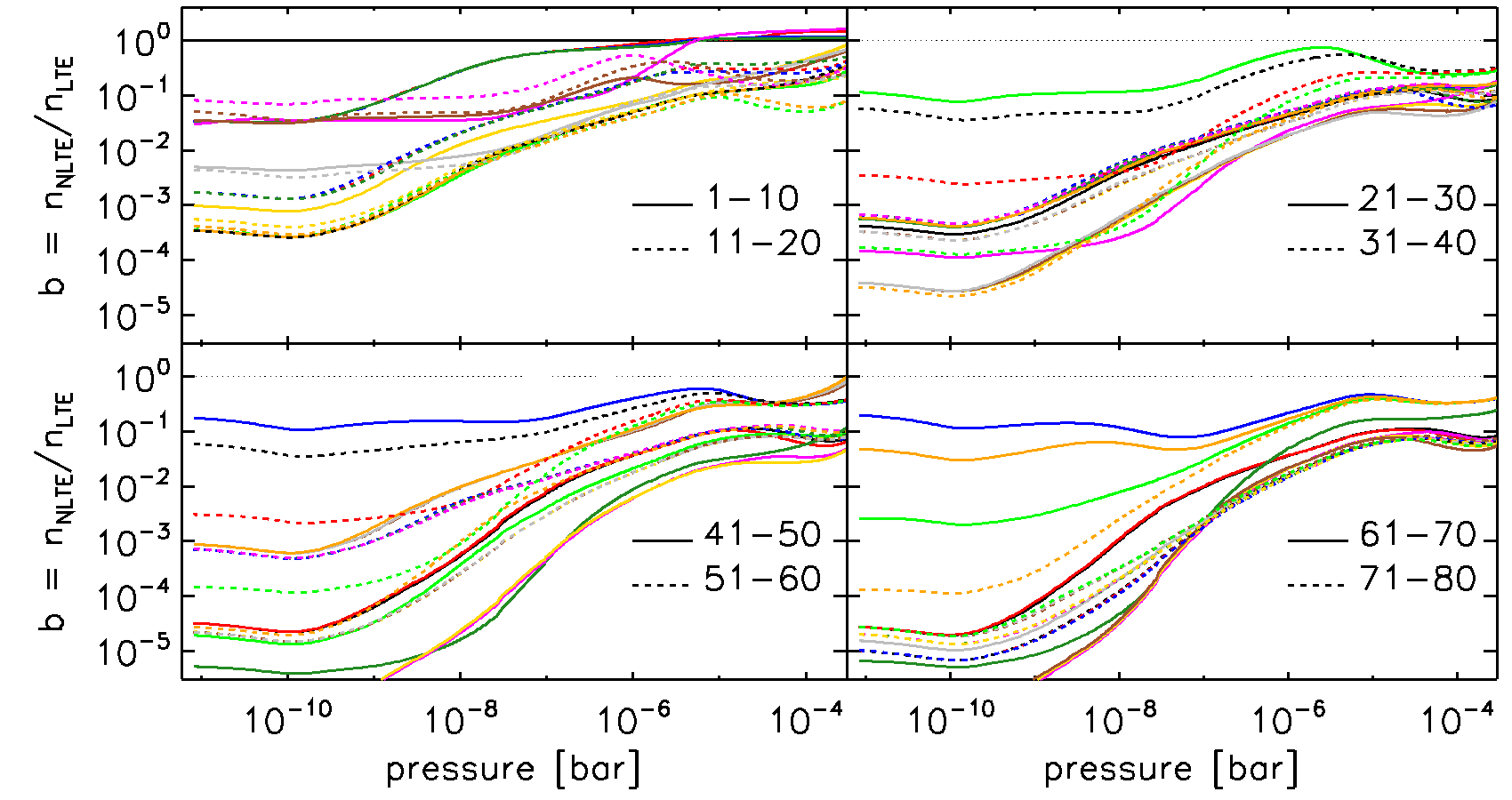}
\caption{Same as Figure~\ref{fig:dep.coeff.Fe1}, but for Mg{\sc i}.}
\label{fig:dep.coeff.Mg1}
\end{figure*}
\begin{figure}[h!]
\includegraphics[width=\hsize,clip]{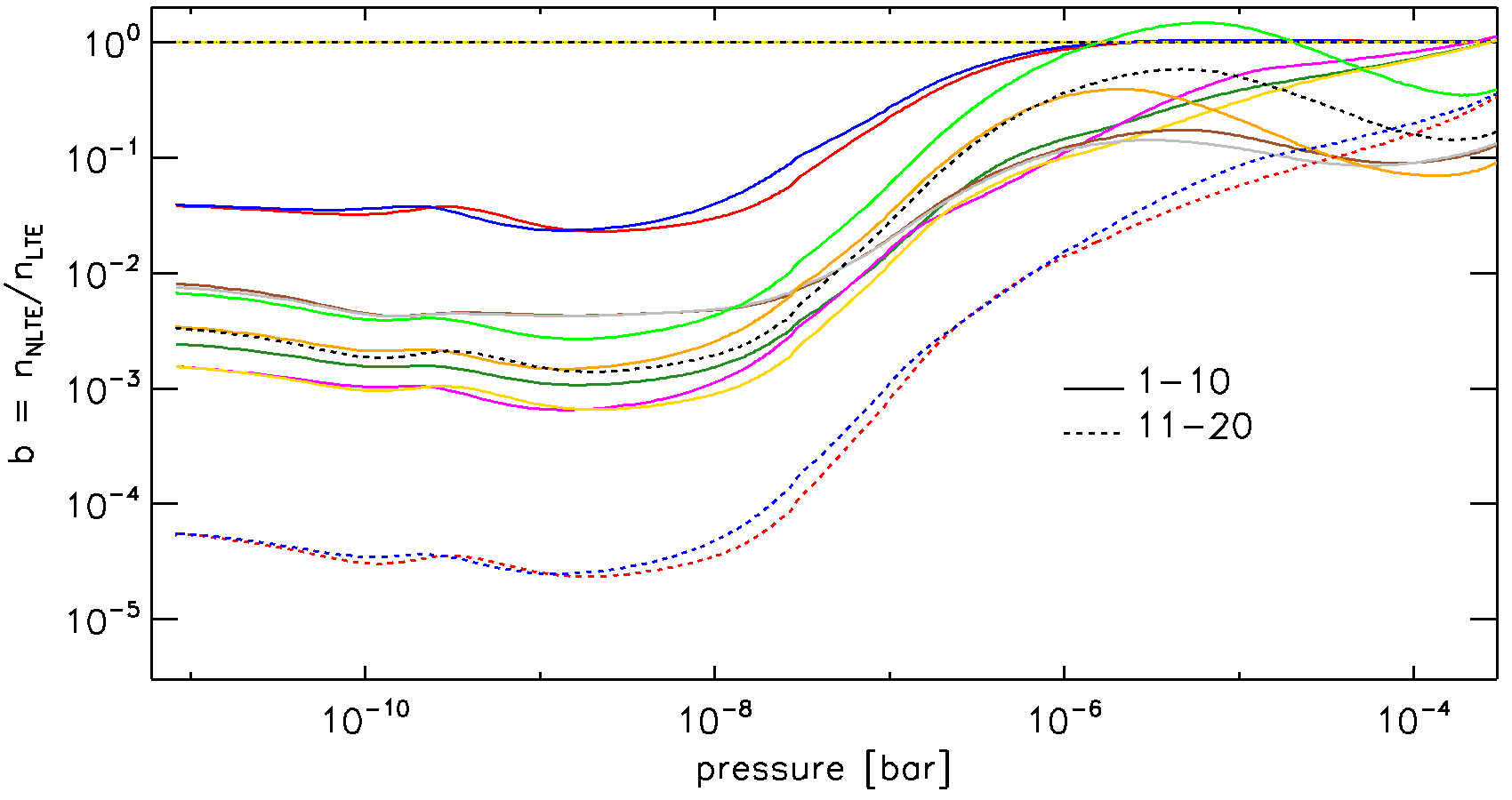}
\caption{Same as Figure~\ref{fig:dep.coeff.Fe1}, but for Mg{\sc ii} and up to the first 20 energy level that are those included in the 17.02 Cloudy distribution.}
\label{fig:dep.coeff.Mg2}
\end{figure}
%
%
\section{Ultraviolet and optical transmission spectra of KELT-9b in NLTE and LTE}\label{sec:appendix_transmissionSpectra}
%
%
\begin{figure*}[h!]
\includegraphics[width=\hsize,clip]{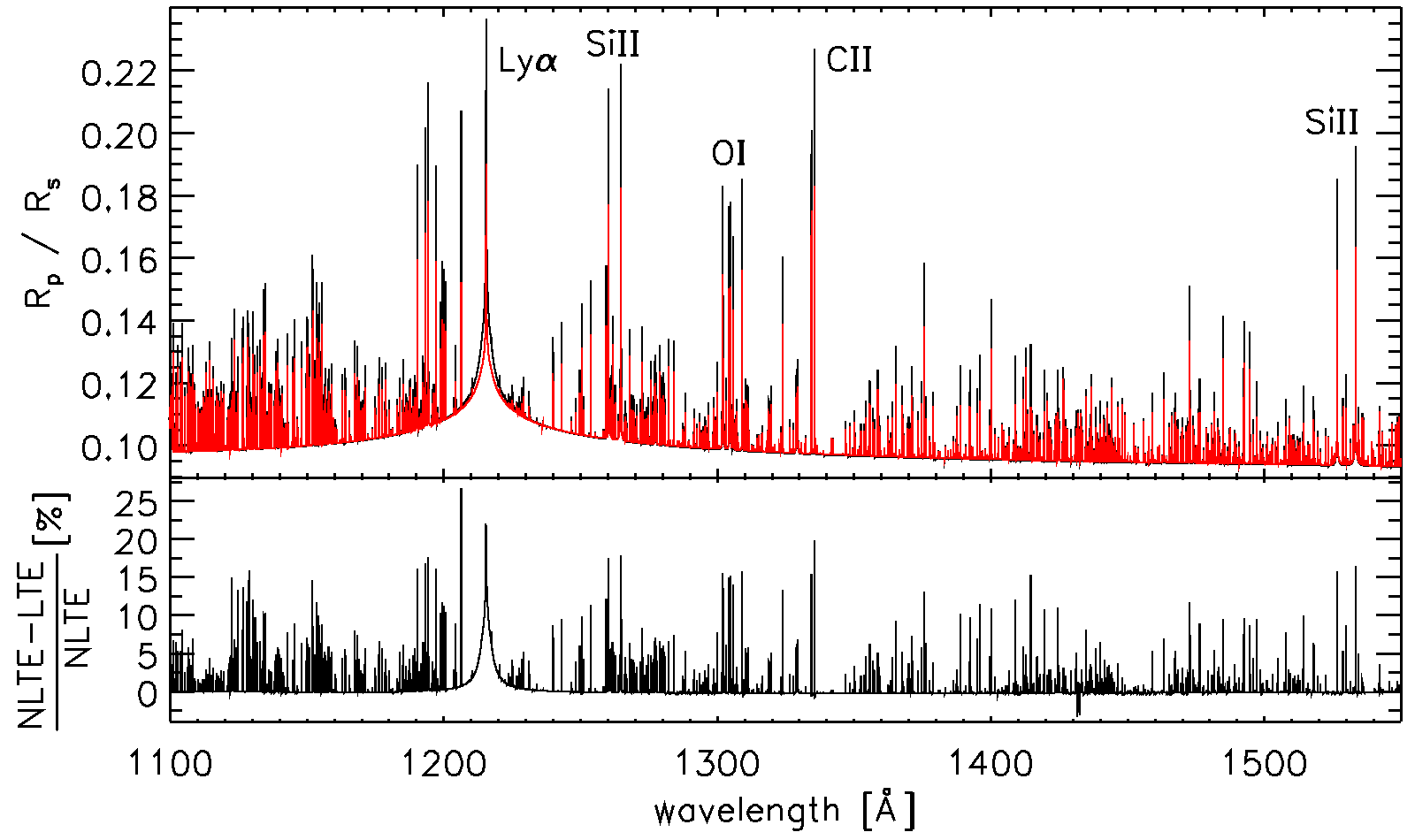}
\includegraphics[width=\hsize,clip]{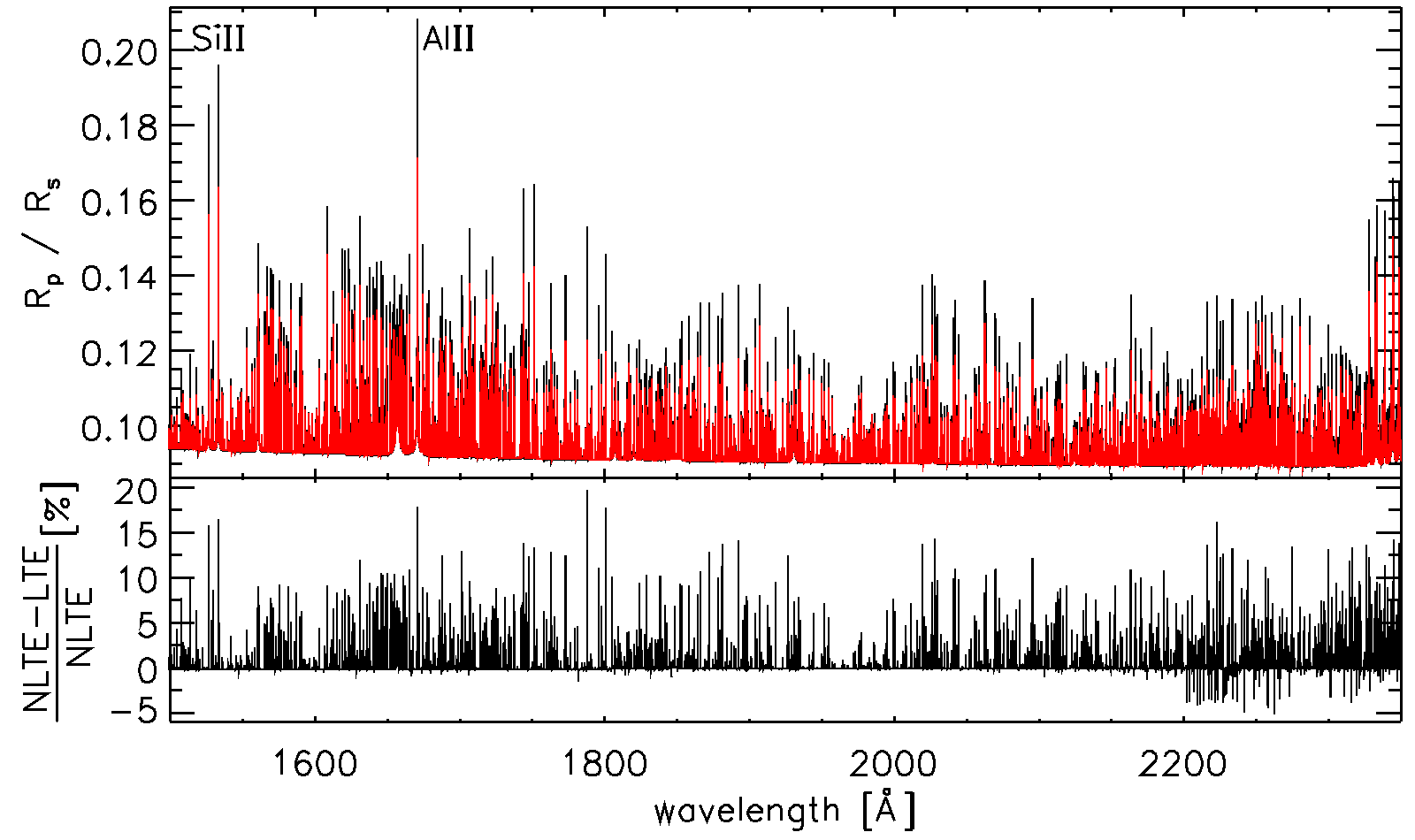}
\caption{Same as Figure~\ref{fig:transmission_all}, but for the 1100--1550\,\AA\ (top) and 1500--2350\,\AA\ (bottom) wavelength ranges.}
\label{fig:transmission_lteVSnlte_ranges1}
\end{figure*}

\begin{figure*}[h!]
\includegraphics[width=\hsize,clip]{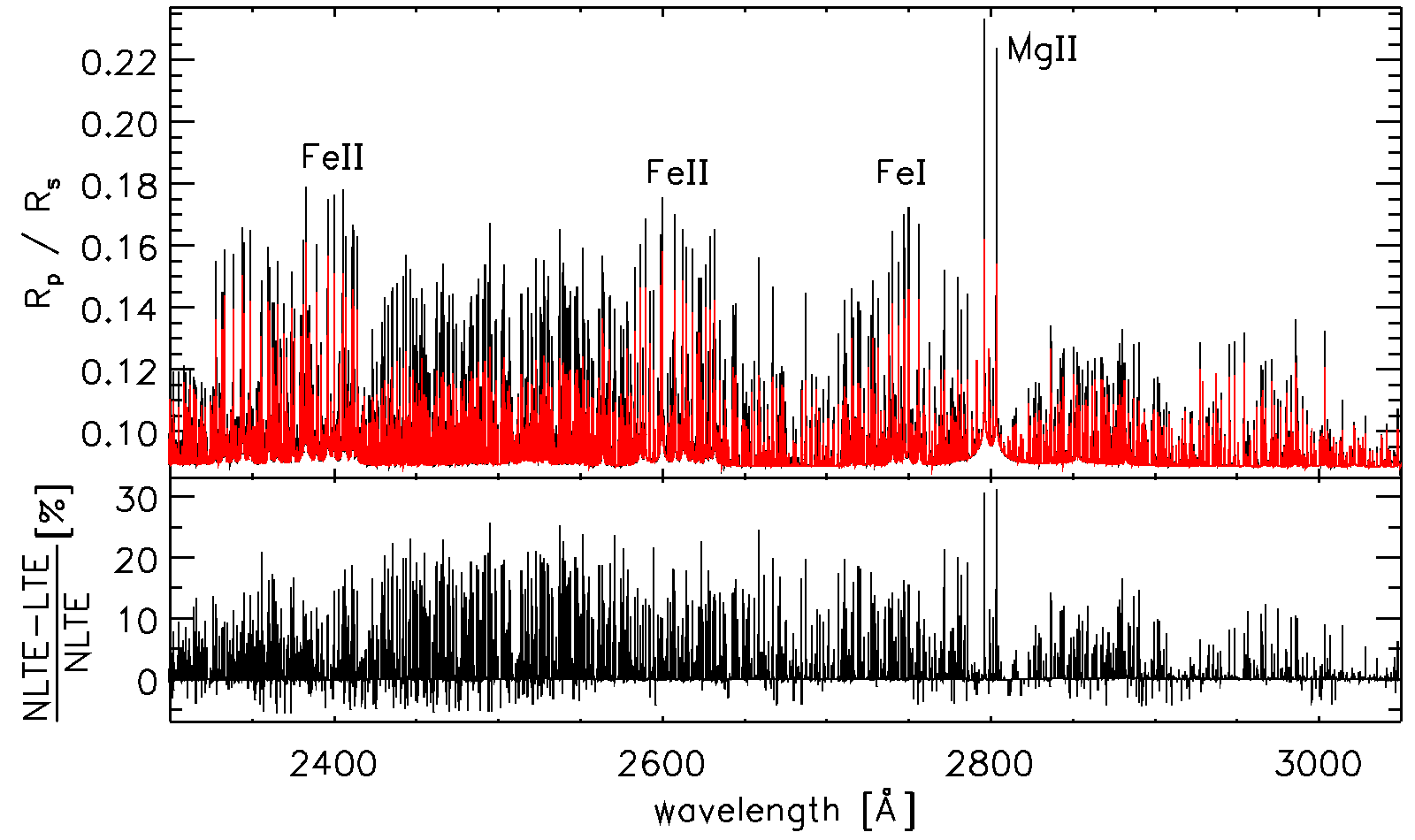}
\includegraphics[width=\hsize,clip]{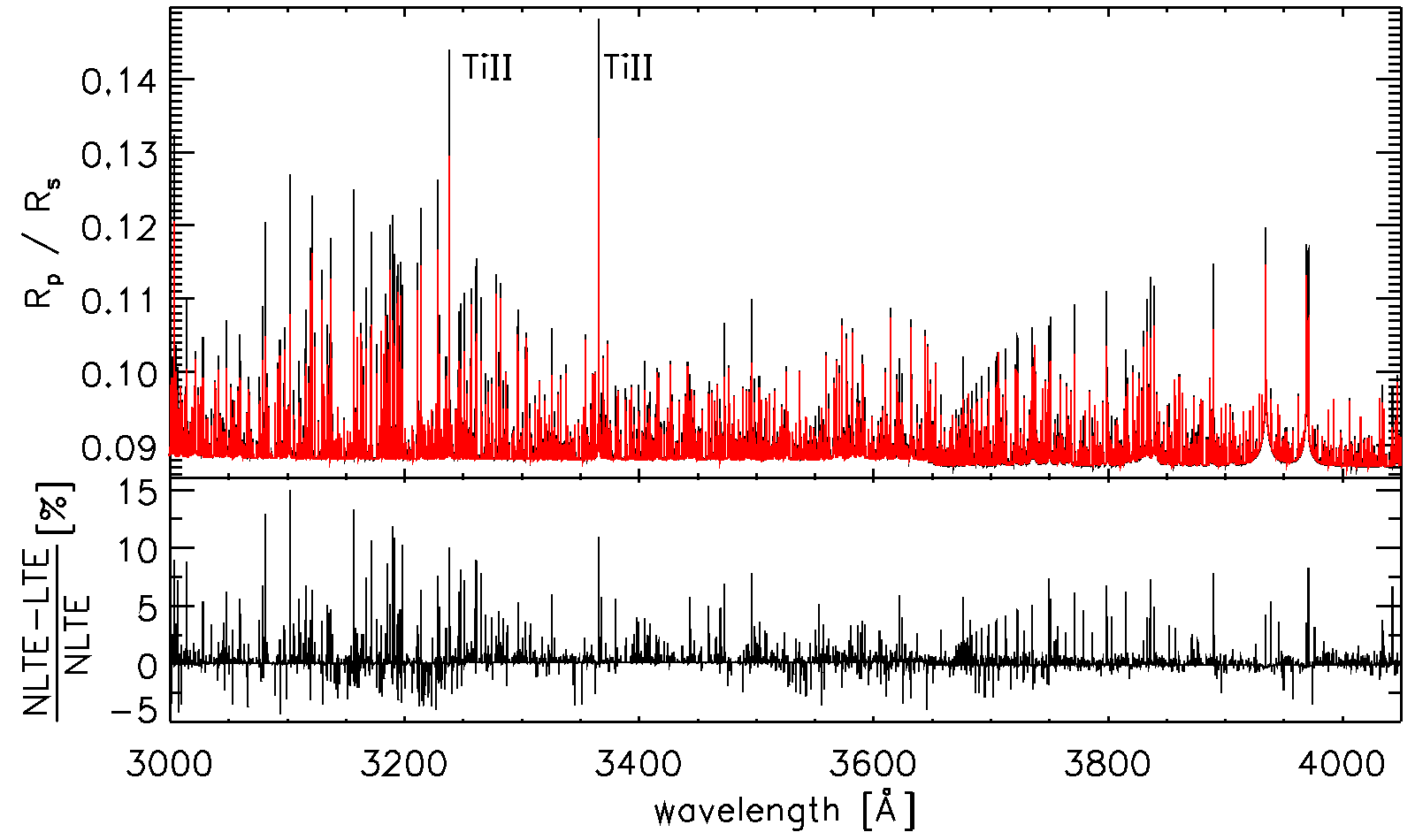}
\caption{Same as Figure~\ref{fig:transmission_all}, but for the 2300--3050\,\AA\ (top) and 3000--4050\,\AA\ (bottom) wavelength ranges.}
\label{fig:transmission_lteVSnlte_ranges2}
\end{figure*}

\begin{figure*}[h!]
\includegraphics[width=\hsize,clip]{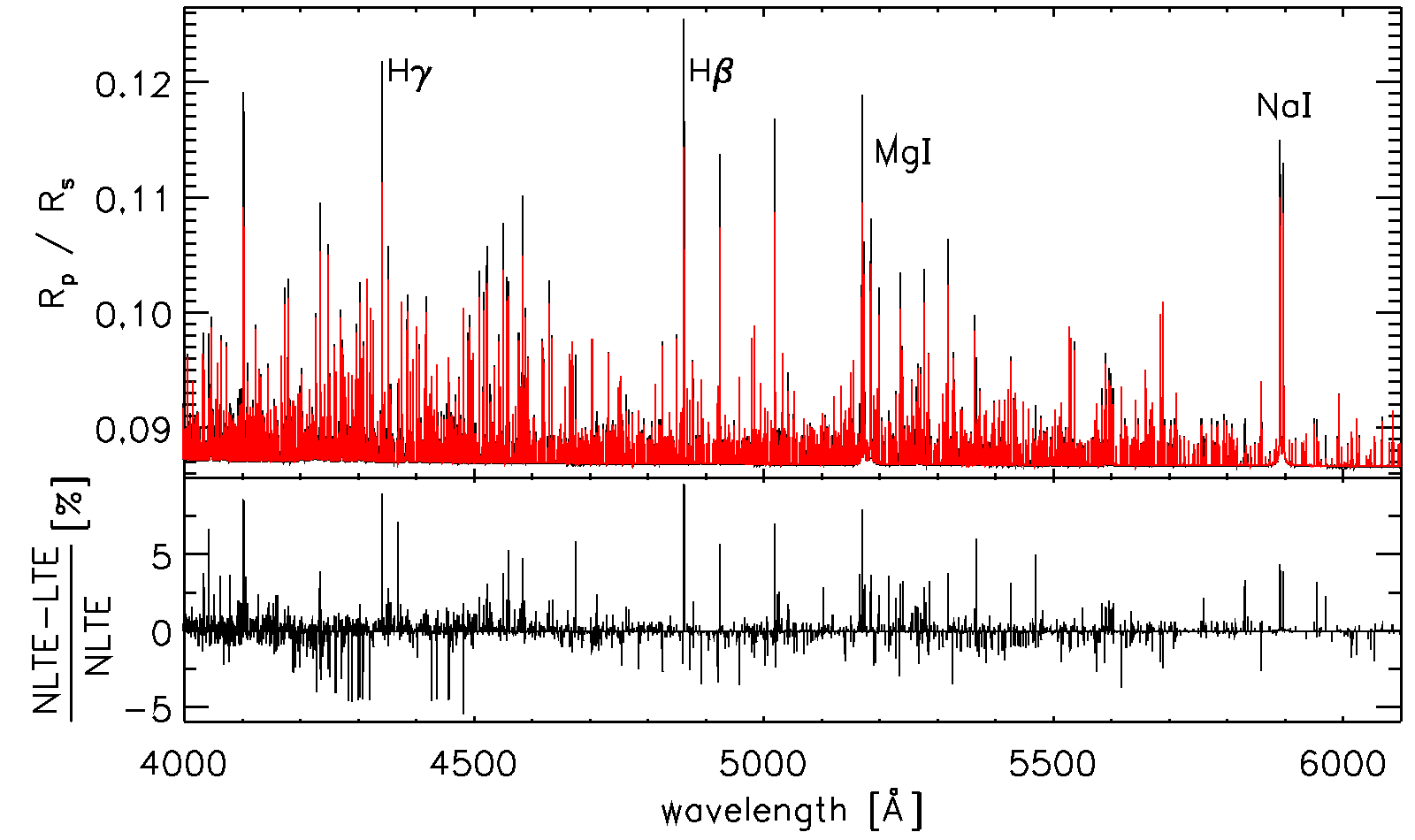}
\includegraphics[width=\hsize,clip]{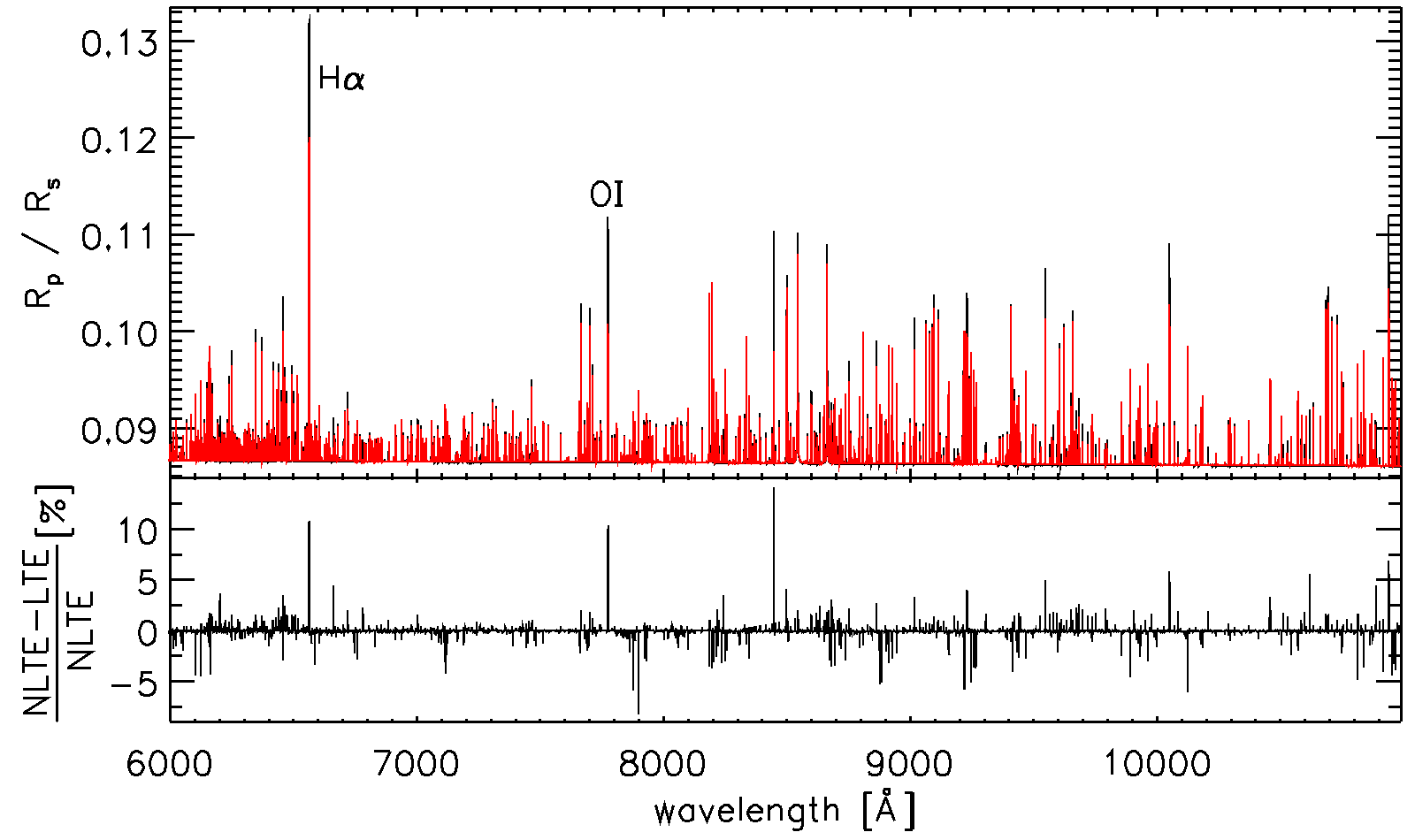}
\caption{Same as Figure~\ref{fig:transmission_all}, but for the 4000--6100\,\AA\ (top) and 6000--11000\,\AA\ (bottom) wavelength ranges.}
\label{fig:transmission_lteVSnlte_ranges3}
\end{figure*}
%
%
\section{Hydrogen departure coefficients}\label{sec:hydrogen_departure_coefficients}
%
\begin{figure}[h!]
\includegraphics[width=\hsize,clip]{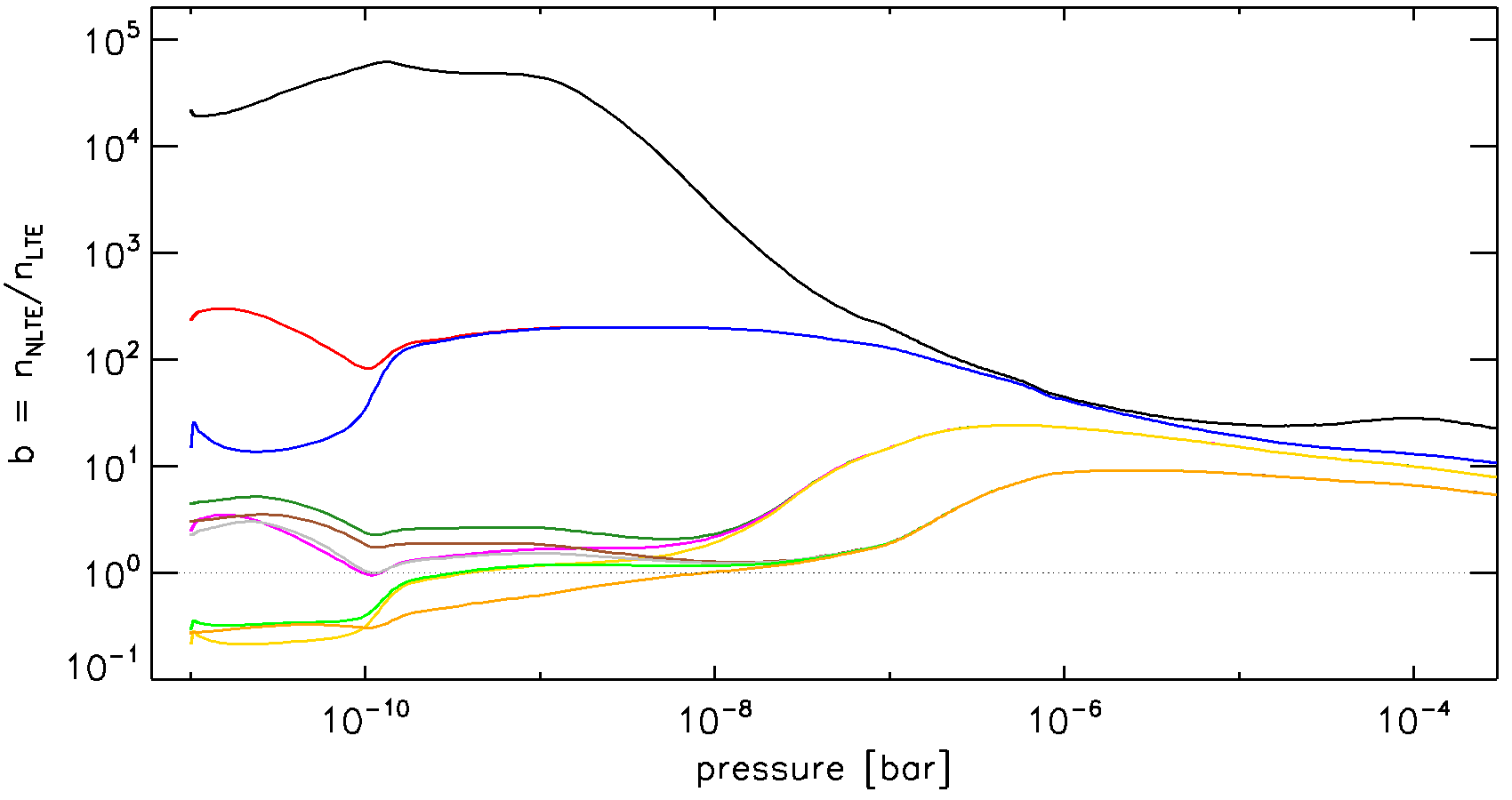}
\caption{Departure coefficients as a function of atmospheric pressure for the first 10 energy levels of H{\sc i}. The order of the line colors corresponding to increasing energy is black, red, blue, dark green, magenta, yellow, brown, gray, bright green, and orange. Therefore, the black solid line shows $b$ for the ground state (i.e. $n$\,=\,1 level), the red solid line shows $b$ for the $n$\,=\,2 level, the blue solid line shows $b$ for the $n$\,=\,3 level, etc.}
\label{fig:hydrogen_b}
\end{figure}
%
%
\section{Simulated transmission spectra obtained following two HST/STIS and ten CUTE transit observations}\label{sec:appendix_observations}
%
%
\begin{figure*}[h!]
\includegraphics[width=\hsize,clip]{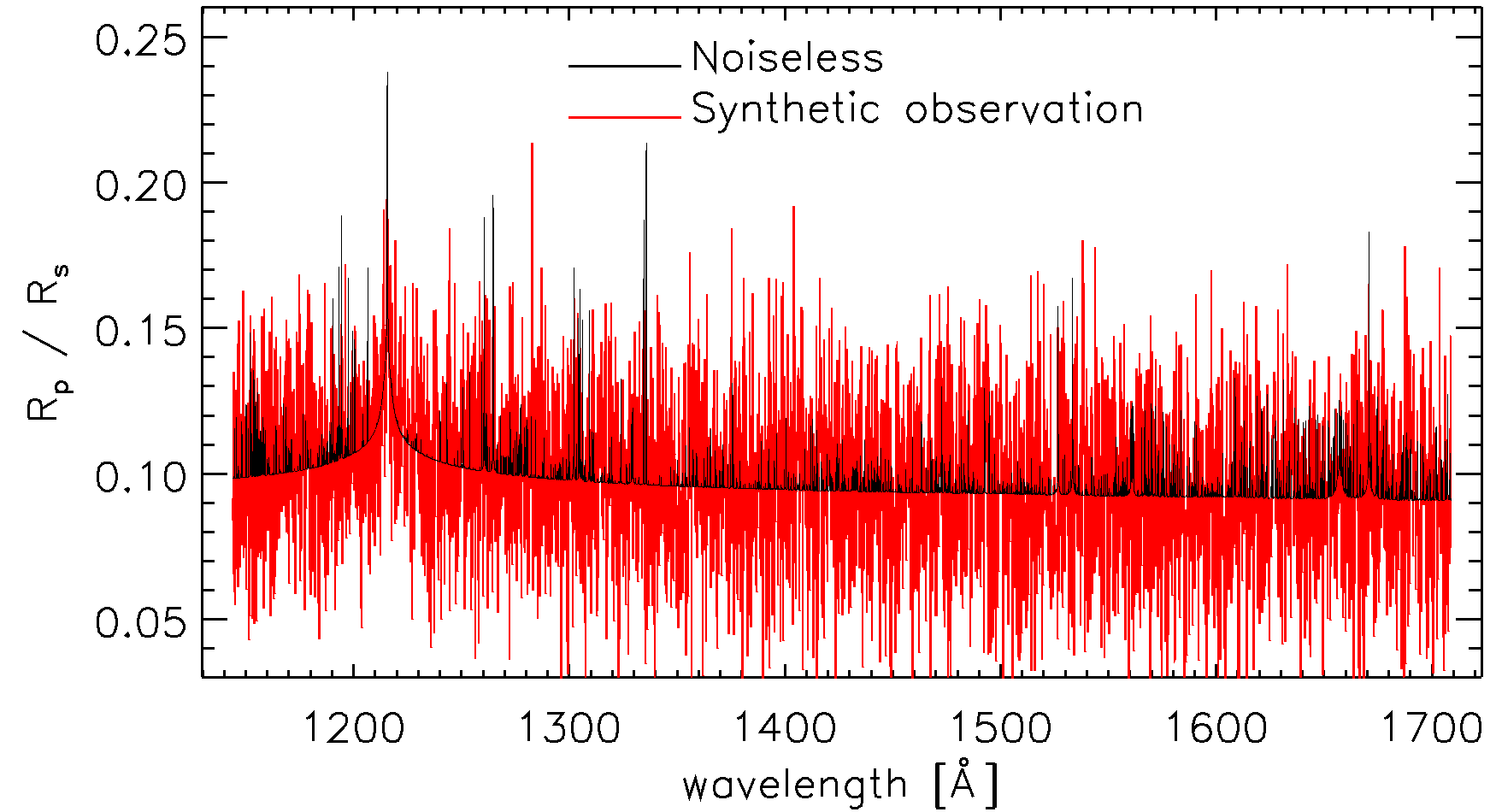}
\includegraphics[width=\hsize,clip]{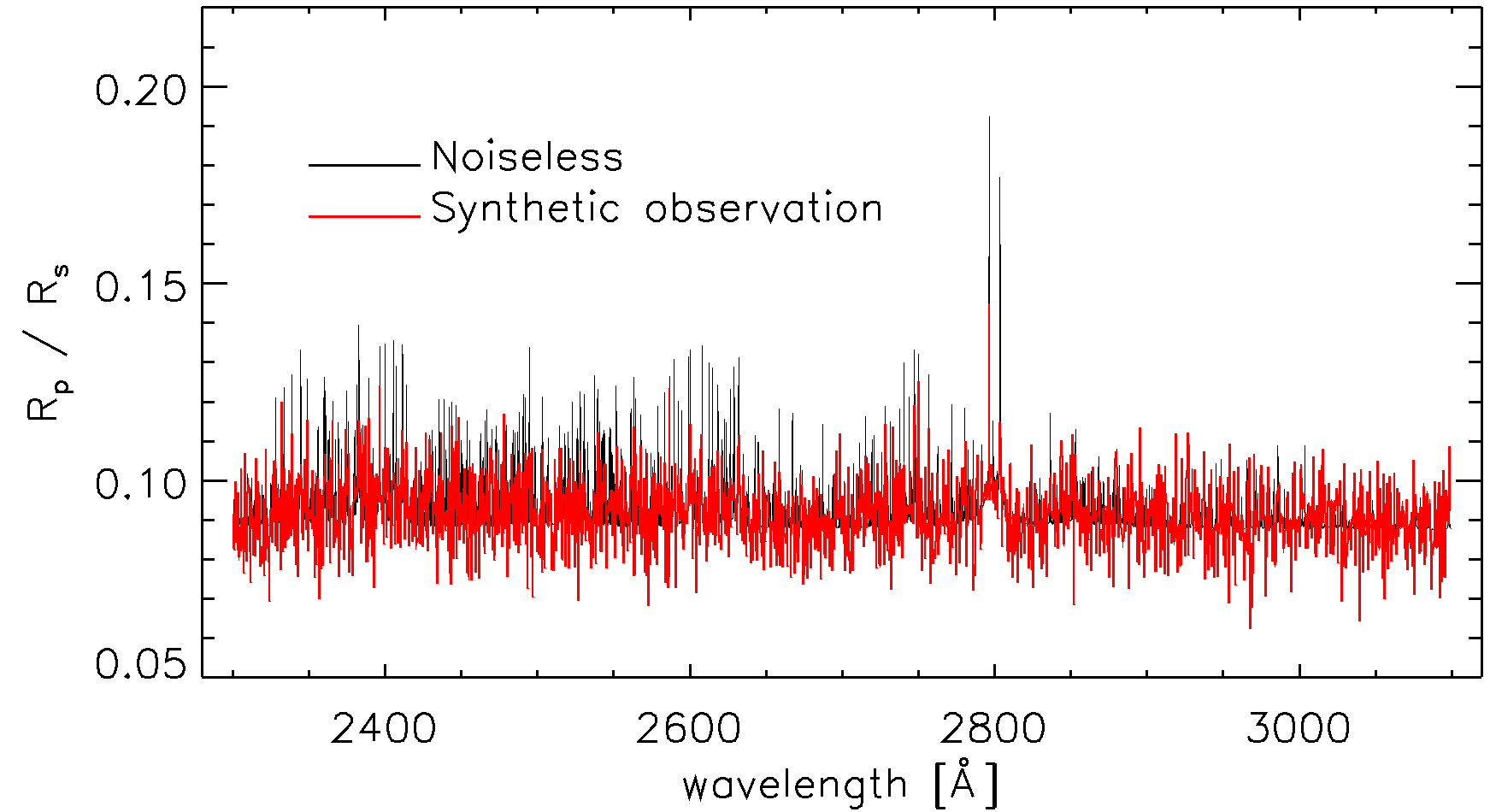}
\caption{Top: simulated transmission spectrum (red line) obtained following two far ultraviolet HST/STIS transit observations collected employing the E140M grating. The transmission spectrum accounts for instrument spectral resolution and wavelength sampling, and it has been obtained binning together ten pixels. The black line shows the transmission spectrum without noise and binning, but accounting for instrument spectral resolution and wavelength sampling. Bottom: same as top, but following two transit observations employing the E230M grating. The black line is underneath the red one for better visibility.}
\label{fig:simulated_observations_HST}
\end{figure*}
%
\begin{figure*}[h!]
\includegraphics[width=\hsize,clip]{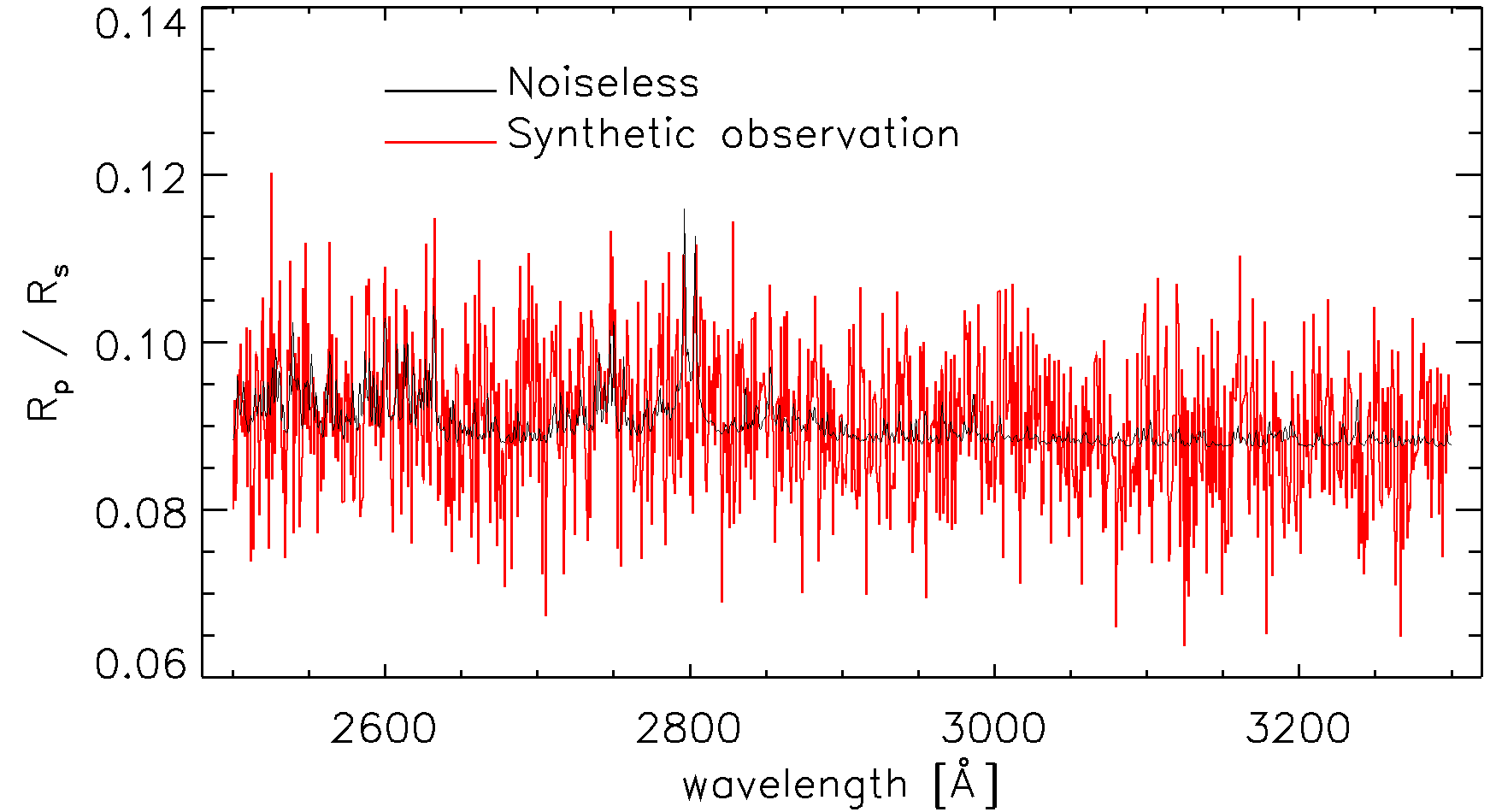}
\caption{Simulated transmission spectrum (red line) obtained following ten CUTE transit observations. The transmission spectrum accounts for instrument spectral resolution and wavelength sampling, and it has been obtained binning together two pixels. The black line shows the transmission spectrum without noise and binning, but accounting for instrument spectral resolution and wavelength sampling.}
\label{fig:simulated_observations_CUTE}
\end{figure*}

%
\end{appendix}
\end{document}